\documentclass[twocolumn, times]{aastex631}
\usepackage{CJK}
\usepackage{amsmath}
\usepackage{subfigure}
\usepackage[T1]{fontenc}

\usepackage{soul}
\setstcolor{red}
\setul{}{1.25pt}

\def\equationautorefname#1#2\null{(#2)}

\begin{document}
\begin{CJK*}{UTF8}{gbsn}

\title{Radiation GRMHD Models of Accretion onto Stellar-Mass Black Holes: I. Survey of Eddington Ratios}

\correspondingauthor{Lizhong Zhang (张力中)}
\email{lizhong4physics@gmail.com}

\author[0000-0003-0232-0879]{Lizhong Zhang (张力中)}
\affiliation{Center for Computational Astrophysics, Flatiron Institute, New York, NY, USA}
\affiliation{School of Natural Sciences, Institute for Advanced Study, Princeton, NJ, USA}

\author[0000-0001-5603-1832]{James M. Stone}
\affiliation{School of Natural Sciences, Institute for Advanced Study, Princeton, NJ, USA}

\author[0000-0003-2131-4634]{Patrick D. Mullen}
\affiliation{Michigan SPARC, Los Alamos National Laboratory, Ann Arbor, MI}
\affiliation{Computational Physics and Methods, Los Alamos National Laboratory, Los Alamos, NM}

\author[0000-0001-7488-4468]{Shane W. Davis}
\affiliation{Department of Astronomy, University of Virginia, Charlottesville, VA, USA}
\affiliation{Virginia Institute for Theoretical Astronomy, University of Virginia, Charlottesville, VA, USA}

\author[0000-0002-2624-3399]{Yan-Fei Jiang (姜燕飞)}
\affiliation{Center for Computational Astrophysics, Flatiron Institute, New York, NY, USA}

\author[0000-0001-7448-4253]{Christopher J. White}
\affiliation{Center for Computational Astrophysics, Flatiron Institute, New York, NY, USA}
\affiliation{Department of Astrophysical Sciences, Princeton University, Princeton, NJ, USA}

\begin{abstract}
We summarize results from a survey of radiation-dominated black hole accretion flows across a wide range of mass accretion rates, as well as two values of black hole spin and initial magnetic field geometry.  These models apply an algorithm targeting direct solutions to the radiation transport equation in full general relativity and have been enabled by access to modern exascale computing systems. Super-Eddington accretion flows form geometrically thick radiation pressure supported disks that drive powerful equatorial outflows.  A narrow funnel-shaped photosphere in the inner region results in very low radiative efficiencies in this regime.  The structure of near- and sub-Eddington accretion depends on whether there is net vertical magnetic flux at the midplane of the disk.  With net flux, the disk forms a thin, dense layer at the midplane surrounded by a magnetically-dominated corona, whereas without net flux the disk remains magnetically dominated everywhere.  Although none of our models achieve the magnetically arrested disk (MAD) regime, those with net vertical flux and a rapidly spinning black hole still produce powerful relativistic jets.  Our calculations adopt simple opacity models (with scalings appropriate to stellar-mass black hole accretion).  We discuss the application of our results to observations of X-ray binaries and ultraluminous X-ray sources such as Cyg X-3 and SS433.  We also speculate on the application of our super-Eddington models to the interpretation of little red dots (LRDs) recently discovered by JWST.  
\end{abstract}


\section{Introduction}
\label{sec:introduction}

Many of the most luminous astrophysical sources are thought to be powered by accretion onto black holes, for example active galactic nuclei (AGN) including quasars \citep{Krolik1999}, X-ray binaries \citep{McClintock2006,Done2007}, ultra-luminous X-ray sources \citep{Kaaret2017,King2023}, gamma-ray bursts \citep{Popham1999}, and tidal disruption events \citep{Gezari2021}.  In these systems the luminosity may be a significant fraction of, or even exceed, the Eddington limit $L_{\mathrm{Edd}} \equiv 4\pi GMc/\kappa_{T}$ for spherical accretion, where $M$ is the mass of the central object and $\kappa_{T}$ is the electron (Thomson) scattering opacity.  Moreover, for a standard $\alpha-$disk model \citep{Shakura1973}, the radiation pressure will {\em exceed} that of the gas within a radius given by
\begin{displaymath}
r/r_{\mathrm{S}} < 170 \alpha^{2/21} (L/L_{\mathrm{Edd}})^{16/21} (M/M_{\odot})^{2/21}
\ ,
\end{displaymath}
where $r_{\mathrm{S}}=2GM/c^2$ is the Schwarzschild radius of the central black hole and $\alpha$ the viscosity parameter.  Thus, the inner regions of all luminous sources with $L/L_{\mathrm{Edd}} \gtrsim 10^{-2}$ should be radiation-dominated, and radiation is therefore an essential ingredient that must be included when modeling the dynamics of such sources. 

Numerical methods have emerged as a powerful tool for studying the physics of general relativistic magnetohydrodynamic (GRMHD) black hole accretion flows.  They enable the assumptions inherent in one-dimensional steady-state self-similar accretion solutions \citep{Shakura1973, Novikov1973, Abramowicz88} to be relaxed, allow study of the multidimensional and time-dependent structure of the disk, and (more importantly) incorporate the effects of outflows on the dynamics self-consistently. More realistic numerical models are motivated by the fact that $\alpha-$disk models do not fit the spectra \citep{DavisTchekhovskoy2020} or size \citep{Morgan2010} of disks in AGN.  They predict radiation-pressure dominated disks should be both viscously and thermally unstable \citep{Lightman1974,Shakura1976}, yet there is little observational evidence for such instabilities.  Some neutron stars and black holes accreting at near- or super-Eddington rates exhibit complex flaring behavior \citep[e.g., ][]{Altamirano2011,Motta2020}, which has been proposed as a manifestation of radiation pressure dominated instability \citep[e.g., ][]{Janiuk2015}, although this connection remains highly uncertain \citep{Blaes2025}.  Finally, such models do not address the production of jets and outflows which are ubiquitous in accretion \citep{Mirabel1999,Blandford2019}. 

Incorporating radiation transfer into numerical methods for modeling the radiation-dominated regime is a significant challenge.  Thus, most previous work has relied upon approximate methods.  For example, early studies used the flux-limited diffusion (FLD) approximation \citep{Levermore1981} to study the properties of MHD turbulence in a radiation-dominated plasma \citep{Turner2003,Turner2004} or to survey the two-dimensional structure of the accretion at different accretion rates \citep{Ohsuga2005,Ohsuga2011}.  In a pioneering set of papers, a two-moment method combined with the M1 approximation \citep{Levermore1984} was implemented in the HARMRAD code \citep{McKinney2014}, and subsequently used to study magnetically arrested disks (MAD) across a range of accretion rates \citep{McKinney2017, Morales-Teixeira2018}. Contemporaneously, the M1 approximation was implemented in the KORAL code \citep{Sadowski2013} and used to study radiation dominated accretion flows including two temperature effects \citep{Sadowski2015,Sadowski2017}.  These early models showed that not only was super-Eddington accretion possible, but also highlighted the important role of magnetic fields in supporting the disk against thermal instability.  More recently, the M1 approximation has been used to compute high-resolution models of radiation dominated accretion flows using the H-AMR code \citep{Liska2022}, as well as to explore the thermal instability and radiative efficiency in thin disks using the Cosmos++ code \citep{Mishra2016, Mishra2022, Fragile2023}.  A review of numerical models of super-Eddington accretion is given in \citet{JiangDai2024}.

Despite the impressive progress in our understanding of radiation dominated accretion enabled by these approximate methods, there remains a concern that detailed aspects of the flow may not be captured correctly.  For example, it is well known that both FLD and the M1 approximation generate incorrect solutions in optically-thin regimes that are likely important in black hole accretion flows \citep{Gonzalez2007, Rosdahl2015, Fragile2020}.  While the general properties of the flow may be unaffected, it is impossible to assess the impact of the failure of these methods on more detailed quantities such as the radiative efficiency, beaming factor, and dynamics of the jets and outflows.  For example, \citet{Asahina2022} has shown that photon collisions inherent in the M1 method artificially enhances the outward radiation flux along the rotation axis in accretion disk simulations.  These issues motivate the use of more accurate methods. 

At low accretion rates, where the radiation pressure is negligible and the primary effect of radiation transport is cooling, more accurate Monte-Carlo methods have been adopted to study accretion onto low-luminosity AGN \citep{Ryan2017, Dexter2021}.  At higher accretion rates, where the radiation pressure becomes important and shot noise in Monte-Carlo methods may become limiting, finite-volume discretizations of the time-dependent transfer equation become advantageous \citep{Jiang2014a}.
In an earlier series of papers these methods were used to study luminous accretion flows onto stellar-mass \citep{Jiang2014b, HuangJiang2023} and supermassive \citep{Jiang2019ApJ880, Jiang2019ApJ885}  black holes at different accretion rates.  This work built upon previous results that used a variable Eddington tensor approach for solving the non-relativistic radiation transport equation \citep{DavisStoneJiang2012, JiangStoneDavis2012} to study the properties of MHD turbulence in the radiation dominated regime \citep{JSD-RadTurb2013}, thermal runaway in AGN disks \citep{JSD-ThermalStab2013, Jiang2016} and the formation of coronae \citep{JSD2014}.  To model the inner regions of black hole accretion flows, however, then clearly GR radiation transport methods are required.

Recently, we have extended our full transport methods for radiation transport to GR \citep{White2023}.  Using a broad range of test problems, we have shown that our methods capture the propagation of photons in curved spacetime accurately, and when coupled to the fluid converges to known solutions for linear waves and shocks.  While the method is significantly more expensive than GRMHD alone (although not much more expensive than approximate transport algorithms such as M1), the emergence of exascale computing platforms has enabled well-resolved parameter surveys of radiation dominated accretion using full transport methods.  In this, the first of a series of papers, we present an overview of our results covering a wide range of mass accretion rates, from sub-Eddington to highly super-Eddington.  In Papers II through IV of this series, we provide a detailed analysis of the structure and dynamics of accretion flows in the super-, near-, and sub-Eddington regimes, respectively, for various black hole spins.  

Because of the complexity of treating opacities in cooler AGN disks, in this series we focus on numerical models that adopt opacities appropriate for accretion onto stellar-mass black holes.  Future work should investigate radiation GRMHD models of accretion onto supermassive black holes using tabular opacities appropriate for this regime.  Additionally, extending the algorithm to incorporate frequency-dependent radiation fields (e.g., \citealt{Fragile2020}; \citealt{Jiang2022}) would facilitate more accurate modeling of sub-Eddington AGN with more realistic frequency-dependent opacities.

The organization of this paper is as follows.  In the next section, we briefly describe our numerical methods, initial and boundary conditions, and the parameters used for the different runs.  In \autoref{sec:results} we describe our main results, followed by a discussion of their observational implications in \autoref{sec:obs_implication}.  We conclude in \autoref{sec:conclusion}.

\section{Numerical Methods}
\label{sec:numerical_method}

Our models are computed using the radiation GRMHD module \citep{White2023} implemented in the \texttt{AthenaK} code \citep{Stone2024}, a new version of \texttt{Athena++} \citep{Stone2020} optimized for exascale computing.  The module solves the coupled conservation laws for mass, energy, momentum, and magnetic flux in a stationary spacetime, along with the time-dependent relativistic radiation transfer equation for the angle-dependent specific intensity.  Energy and momentum exchange between the radiation and matter is calculated directly from angular quadratures.  The full system of equations solved is: 
\begin{subequations}    
\begin{align}
    & \partial_0(\rho u^0) + \partial_j(\rho u^j) = 0
    \ ,
    \\
    & \partial_0(T^0_{\ \alpha}) + \partial_j(T^j_{\ \alpha}) = T^{\beta}_{\ \lambda}\Gamma_{\alpha\beta}^{\lambda} + G_{\alpha}
    \ ,
    \label{eq:stress_energy}
    \\
    & \partial_0(B^i) + \partial_j(b^i u^j - b^j u^i) = 0
    \ ,
    \\
    & \begin{aligned}
    & \partial_0(\hat{n}^0\hat{n}_0\hat{I}) + \partial_j(\hat{n}^j\hat{n}_0\hat{I}) + \frac{1}{\sin\hat{\zeta}}\partial_{\hat{\zeta}}(\sin\hat{\zeta}\hat{n}^{\hat{\zeta}}\hat{n}_0\hat{I}) 
    \\
    &\mkern+160mu + \partial_{\hat{\psi}}(\hat{n}^{\hat{\psi}}\hat{n}_0\hat{I})= \hat{n}_0 (\hat{j} - \hat{\chi}\hat{I}) 
    \ , 
    \label{eq:rad_transfer}
    \end{aligned}        
\end{align}
\end{subequations}
where the primitive variables gas density $\rho$, gas pressure $P_g$, four-velocity $u^{\alpha}$, and the three- and four-magnetic field $B^i$ and $b^\alpha$ are defined in the normal frame.  We adopt Cartesian Kerr-Schild coordinates where $\sqrt{-g}=1$, where $g= \mathrm{det} \; \mathbf{g}$ and $\mathbf{g}$ is the metric.

The radiation transfer equation~\autoref{eq:rad_transfer} is frequency-integrated and written in flux-conservative form \citep{Davis2020} in the tetrad frame (denoted with circumflexes).  This enables adoption of mature and accurate numerical methods developed for solving hyperbolic conservation laws in general.  The frequency-integrated specific intensity $\hat{I}$ is discretized in angle on a geodesic grid, with direction $\hat{n}^{\alpha}$ determined by the polar angle $\hat{\zeta}$, azimuthal angle $\hat{\psi}$, and the local tetrad.  The definitions of $\hat{n}^{\hat{\zeta}}$ and $\hat{n}^{\hat{\psi}}$ in the angular fluxes can be found in equation~(8) of \citet{Davis2020} or equation~(16) of \citet{White2023}.  

The geometric source terms related to the connection coefficients $\Gamma_{\alpha\beta}^{\lambda}$ are introduced by the covariant derivatives in curved spacetime.  To simplify these terms we follow the convention used in the HARM code \citep{Gammie2003} and write the stress-energy tensor $T^{\alpha}_{\ \beta}$ in mixed form as
\begin{equation}
\begin{aligned}
    & T^{\alpha}_{\ \beta} = \left(\rho + \frac{\gamma}{\gamma-1} P_g + b^{\lambda}b_{\lambda}\right) u^{\alpha}u_{\beta} 
    \\
    &\mkern+160mu + \left(P_g+\frac{1}{2}b^{\lambda}b_{\lambda}\right) \delta^{\alpha}_{\ \beta} - b^{\alpha}b_{\beta}
    \ , 
\end{aligned}
\end{equation}
where $\delta^{\alpha}_{\ \beta}$ is the Kronecker delta. For the equation of state, we adopt an ideal gas law with adiabatic index $\gamma=5/3$.  Note this differs from the value $\gamma=4/3$ usually adopted for GRMHD models of black hole accretion flows \citep{porth2019, Gammie2025}. 

The full details of the numerical algorithms used to integrate these equations are described in \citet{White2023} and \cite{Stone2024} and will not be repeated here. However, it is important to note that from the various algorithmic options available in \texttt{AthenaK}, we adopt the second-order Runge-Kutta time integrator with a Courant-Friedrichs-Lewy number of 0.3 and with fluxes computed using the HLLE Riemann solver.  Spatial reconstruction in the GRMHD and radiative transfer equations use the PPM4 and PLM algorithms, respectively.  Angular fluxes in the radiation transfer equation are computed using donor cell reconstruction.

\subsection{Radiation Source Terms and Opacity}

The radiation source terms, characterized by the emissivity $\hat{j}$ and absorption coefficient $\hat{\chi}$, include emission, absorption, and scattering processes and are solved in the fluid frame.  The radiation feedback is then evaluated through the coupling term
\begin{equation}
    G_{\alpha} = -\oint \hat{n}_{\alpha}(\hat{j}-\hat{\chi}\hat{I})d\hat{\Omega}
    \ , 
\end{equation}
where $d\hat{\Omega}=\sin\hat{\zeta}d\hat{\zeta}d\hat{\psi}$ is the differential solid angle in the tetrad frame.  

The radiation source terms in the fluid frame (denoted by bars) are defined as follows: 
\begin{subequations}
\begin{align}
    &\begin{aligned}
    \bar{j} &= \rho\kappa_P \left(\frac{a_r T_g^4}{4\pi} - \bar{J}\right) + \rho(\kappa_R+\kappa_T)\bar{J}         
    \\
    &\mkern+180mu + \rho\kappa_T\frac{4k_B(T_g-T_r)}{m_e c^2}\bar{J}
    \label{eq:rad_emission}
    \ ,
    \end{aligned}
    \\
    &\bar{\chi} = \rho(\kappa_R+\kappa_T)
    \label{eq:rad_absorption}
    \ ,
\end{align}
\end{subequations}
where $\bar{J}=(4\pi)^{-1}\oint\bar{I}d\bar{\Omega}$ is the zeroth angular moment of the fluid-frame intensity $\bar{I}$, integrated over the fluid-frame solid angle $\bar{\Omega}$.  For an ideal gas law, the gas temperature is $T_g=P_g \bar{m}/(\rho k_B)$, where $k_B$ is the Boltzmann constant and $\bar{m}=0.5m_p$ represents the mean molecular weight of ionized hydrogen. The radiation effective temperature is $T_r=(4\pi\bar{J}/a_r)^{1/4}$.

A crucial ingredient of radiation GRMHD models is the use of accurate opacities. The models computed in this paper use frequency-integrated opacities appropriate for accretion onto
stellar-mass black holes.  At the high temperatures characteristic of such flows ($T_g \gtrsim 10^{7}$~K), the only significant sources of opacity are electron scattering and free-free absorption.  In this work, we adopt the Thomson opacity $\kappa_T=0.4~\mathrm{cm^2~g^{-1}}$ for electron scattering, and use Kramers' opacity for the Planck mean emission $\kappa_P=1.55\times10^{24} \rho T_g^{-7/2}$~$\mathrm{cm^2~g^{-1}}$, and the Rosseland mean absorption $\kappa_R=4.18\times10^{22} \rho T_g^{-7/2}$~$\mathrm{cm^2~g^{-1}}$, respectively.  At the lower temperatures expected in AGN accretion flows around supermassive black holes, more complex opacities \citep{Iglesias1996} are required. For example, the iron opacity bump can change the dynamics of the disk significantly \citep{Jiang2016,Jiang2020}.  In this case, the use of simple Rosseland and Planck means for the opacities may be unjustified, and a full-frequency dependent transport model that directly computes the appropriate flux and emission means may be required.  For these reasons, we focus solely on accretion models onto stellar-mass black holes in this paper.

Compton scattering is an important process that couples the gas and radiation field at high temperatures.  We approximate the energy exchange via Compton scattering using the last term in \autoref{eq:rad_emission} based on the methods used in \citep{Blaes2003, Hirose2009}.  More accurate methods for Compton scattering have been introduced by \citep{Sadowski2015b, McKinney2017}. An even more accurate treatment requires solving the frequency-dependent transfer equation \citep{Jiang2022}. Compton scattering is particularly important for capturing the temperature in the optically thin corona that forms above geometrically thin disks in the sub-Eddington regime.  This provides further motivation for future calculations of frequency-dependent models of sub-Eddington disks.

Finally, in the very low density upper regions of the disk, thermal coupling between ions and electrons may be inefficient. A variety of previous studies have considered two-temperature radiation GRMHD models of disks \citep{Sadowski2017, Liska2022}.  However, such calculations rely on uncertain prescriptions for electron and ion heating in a turbulent, radiation-dominated plasma.  Moreover, as discussed above, there are additional uncertainties related to the proper treatment of mean opacities and Compton scattering that also affect the accuracy of the resulting solutions.  For these reasons we restrict our models to single-temperature plasmas.  Investigating kinetic effects in the low-density coronal regions is another important direction for future work.

\subsection{Initial and Boundary Conditions}

\begin{figure*}
    \centering
    \includegraphics[width=\textwidth]{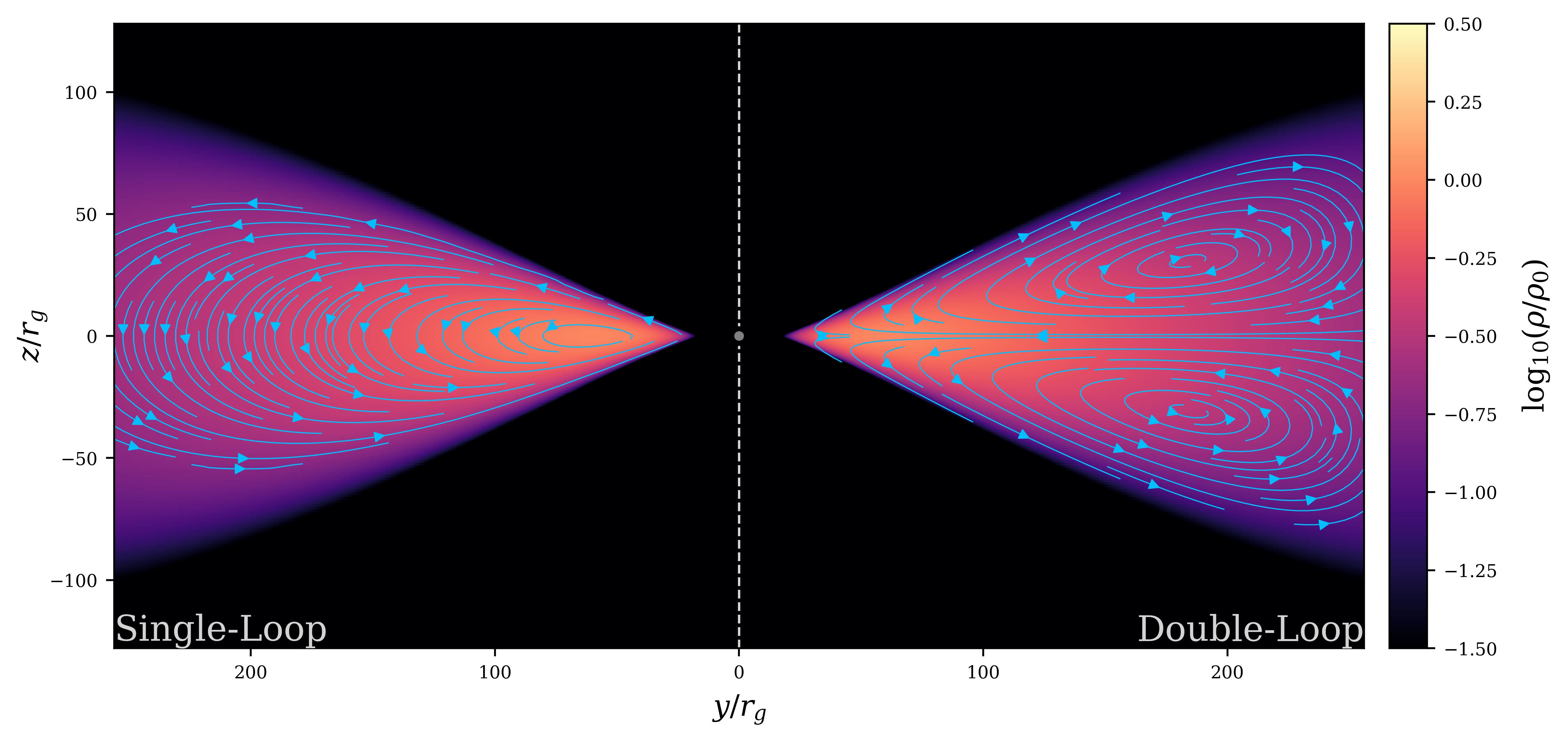}
    \caption{Density (colors) and magnetic field (blue streamlines) in the initial conditions for the single-loop (left) and double-loop (right) magnetic configurations.  }
    \label{fig:ini}
\end{figure*}

Most GRMHD models of black hole accretion flows begin with rotating torus with constant angular momentum in hydrostatic equilibrium \citep{FM1976}.  However, when the torus spans a large range of radii, the assumption of constant angular momentum leads to a very hot, geometrically thick structure.  We have found that in the radiation-dominated regime, such tori undergo rapid evolution due to cooling, and the outgoing radiation field during the subsequent evolution is dominated by cooling from the disk rather than energy released by accretion.  Thus, in this work, our simulations are initialized with a much cooler and thinner disk with an angular momentum profile that increases with radius \citep{Chakrabarti1985}.

The initial torus is configured with an inner edge at $15r_g$, where $r_g = GM/c^2$ is the gravitational radius.  The density and pressure peak at $58r_g$, with both profiles following power laws with indices of $-1.5$ and $-2.5$, respectively.  The peak density is set to an arbitrary value $\rho_0$ that allows the resulting accretion flow to achieve different Eddington ratios for the specified black hole mass. The background medium is at rest with a uniform density $10^{-5}\rho_0$ and pressure $3.3\times10^{-8}\rho_0 c^2$. The equilibrium state is calculated using the total thermal pressure (gas + radiation) assuming thermal equilibrium ($T_g = T_r$). 

A weak magnetic field is added to the torus with either a single-loop or double-loop geometry, initialized using a purely azimuthal vector potential as defined by equation~(2) in \citet{White2020}. The field strength is normalized such that the magnetic pressure is initially 1\% of total thermal pressure. The geometry of the magnetic field in our two initial configurations is shown in \autoref{fig:ini} plotted over the density profile in the torus. Note that the single-loop configuration has a net poloidal flux at the midplane, while the double-loop does not.  This can result in different evolutionary outcomes.

Our simulations use a Cartesian grid.  Within the horizon region, we perform excision (similar to \citealt{Ressler2023}) by setting the density and pressure to floor values (see below), the velocity to free-fall, and the radiation field to zero. However, the magnetic field is still allowed to freely evolve.  

At the outer edges of the domain the boundary conditions for the gas, radiation, and magnetic field are set to outflow.  

To improve the stability of the numerical integration, we use the first-order flux correction algorithm of \citet{Lemaster2009} in cells where the nonlinear conversion from conserved to primitive variables fails.  
This allows us to use much lower floors near the horizon than would otherwise be required: we apply density and pressure floors of $10^{-7}\rho_0$ and $3.33\times10^{-13}\rho_0 c^2$, respectively, uniformly throughout the domain. 
We also apply the first-order flux correction to the two cells adjacent to the horizon.  

To be certain that a radiation-dominated accretion flow is established self-consistently, we evolve the torus using the full radiation GRMHD algorithm from the initial conditions, rather than using an artificial cooling function at early times.

\subsection{Mesh and Resolution}
\label{sec:mesh_and_resolution}

The spatial mesh used in all our calculations employs a Cartesian grid with a domain size of $2048r_g$ in each dimension with increasing levels of static mesh refinement towards the central black hole.  We adopt three grid configurations in our simulations.  For the low-resolution configuration, the root grid consists of $128^3$ cells of size $16r_g$, with 8 levels of refinement applied hierarchically with the most refined region spanning $-8<x,y<8$ and $-4<z<4$.  In the intermediate-resolution configuration, the root grid has $256^3$ cells of size $8r_g$ and 7 levels of refinement, with the most refined region spanning $-16<x,y<16$ and $-8<z<8$.  The high-resolution configuration is identical to the intermediate-resolution setup but adds an 8th level of refinement near the midplane, covering $-28<x,y<28$ and $-2<z<2$.  Both low- and intermediate-resolution simulations achieve the same highest resolution near the black hole, with a cell size of $0.0625r_g$.  However, the intermediate-resolution grid has twice the resolution throughout the rest of the domain compared to the low-resolution grid.  The high-resolution grid is specifically designed to resolve thin disks, with the highest resolution at the midplane region giving a cell size of $0.03125r_g$.

The angular mesh used to discretize the frequency-integrated intensity $\hat{I}$ is based on a level-2 geodesic grid consisting of 42 angles in every spatial grid cell. Figure~3 in \citet{White2023} shows the geometry of this mesh.

The high- and intermediate-resolution simulations were conducted on Frontier at the OLCF (with four AMD MI250X GPUs per node), each requiring approximately 120,000 and 21,000 node hours respectively to reach $t=60000r_g/c$.  The low-resolution simulations were performed on Polaris (with four NVIDIA A100 GPUs per node), each taking only around 4,500 node hours to reach the same simulation time.  Details on the algorithm's scaling performance can be found in \citet{White2023}.

\begin{deluxetable*}{lccccccccccccccc}
\tablecaption{Comparison of time- and azimuthally-averaged properties of the models \label{tab:sim_result}}
\tablehead{
\colhead{Name} & \colhead{$\rho_0$} & \colhead{Resolution} & 
\colhead{$\big<\dot{M}_3\big>_t$} & \colhead{$\left<\varphi_3\right>_t$} & 
\colhead{$H_{10}$} & \colhead{$H_{10}^{\mathrm{(mag)}}$} & 
\colhead{$\tilde{T}_g^{\mathrm{(disk)}}$} & 
\colhead{$\eta_{20}^{\mathrm{(rad)}}$} & 
\colhead{$\eta_{200}^{\mathrm{(wind)}}$} & 
\colhead{$\eta_{20}^{\mathrm{(jet)}}$} & 
\colhead{max$\left(u^t_{\mathrm{(jet)}}\right)$} & 
\colhead{$\big<\dot{M}_{1024}^{(\mathrm{out})}\big>_t$}
\\
& \colhead{$\left(\mathrm{g/cm^3}\right)$} & & 
\colhead{$\left(\dot{M}_{\mathrm{Edd}}\right)$} & & 
\colhead{$\left(r_g\right)$} & \colhead{$\left(r_g\right)$} & \colhead{$\left(10^7~\mathrm{K}\right)$} & 
\colhead{(\%)} & \colhead{(\%)} & 
\colhead{(\%)} & & 
\colhead{$\left(\big<\dot{M}_3\big>_t\right)$}
\\
\quad\;\;\;(1) & (2) & (3) & (4) & (5) & 
(6) & (7) & (8) & (9) & (10) & 
(11) & (12) & (13)
}
\startdata
    E150-a9        & 1e-1 & Med &      148.88 & 1.85 & 2.49 & 5.00 & $6.73~r_{10}^{-0.24}$ & 0.45 & 0.32 & 1.50 &    10.19 & -1.03 \\
    \;\:E88-a3     & 3e-2 & Med &  \;\: 88.77 & 3.12 & 2.09 & 4.34 & $5.46~r_{10}^{-0.18}$ & 0.53 & 0.25 & 0.59 & \;\:8.91 & -0.76 \\
    \;\:E31-a3-DL  & 8e-4 & Low &  \;\: 30.56 & 1.24 & 2.12 & 4.23 & $3.85~r_{10}^{-0.25}$ & 0.49 & 0.25 & ---  & \;\:---  & -2.16 \\
    \;\:E15-a9     & 1e-2 & Med &  \;\: 14.92 & 1.83 & 2.24 & 4.50 & $3.84~r_{10}^{-0.25}$ & 1.63 & 0.59 & 1.75 & \;\:9.54 & -1.09 \\
    \;\:\;\:E9-a3  & 3e-3 & Med &  \quad 9.30 & 2.42 & 1.81 & 3.50 & $3.22~r_{10}^{-0.14}$ & 1.72 & 0.43 & 0.53 & \;\:8.66 & -1.11 \\
    \;\:E09-a3-DL  & 3e-5 & Low &  \quad 0.91 & 1.46 & 1.22 & 1.92 & $1.29~r_{10}^{-0.39}$ & 2.16 & 0.18 & ---  & \;\:---  & -0.16 \\
    \;\:E07-a3-DL  & 8e-5 & Med &  \quad 0.73 & 0.77 & 0.35 & 0.70 & $1.63~r_{10}^{-0.05}$ & 2.44 & 0.85 & ---  & \;\:---  & -2.66 \\
    \;\:E08-a9     & 2e-3 & Med &  \quad 0.80 & 1.75 & 0.22 & 2.56 & $1.70~r_{10}^{-0.24}$ & 5.40 & 2.37 & 0.51 & \;\:4.68 & -2.64 \\
    \;\:E08-a3     & 8e-4 & Med &  \quad 0.84 & 2.84 & 0.24 & 2.63 & $1.49~r_{10}^{-0.38}$ & 3.87 & 1.60 & 0.55 & \;\:6.17 & -4.50 \\
    \;\:E01-a3     & 1e-4 & High & \quad 0.12 & 2.68 & 0.20 & 2.37 & $0.86~r_{10}^{-0.33}$ & 5.41 & 1.26 & 0.20 & \;\:3.17 & -2.17 \\
    \hline
\enddata
\tablecomments{
    {\bf Columns (from left to right):}  
    (1) Model name (see naming convention in \autoref{sec:parameter_values}); 
    (2) Simulation density unit; 
    (3) Resolution (see text in \autoref{sec:mesh_and_resolution}); 
    (4) Accretion rate at $r=3r_g$; 
    (5) Normalized magnetic flux at $r=3r_g$; 
    (6) Density scale height at $r=10r_g$; 
    (7) Magnetic scale height at $r=10r_g$; 
    (8) Power-law fitted gas temperature of the disk body; 
    (9) Radiation efficiency at $r=20r_g$; 
    (10) Wind efficiency at $r=200r_g$; 
    (11) Jet efficiency at $r=20r_g$; 
    (12) Maximum Lorentz factor in jet; 
    (13) Mass loss rate at $r=1024r_g$, expressed in units of the inner accretion rate.  
}
\end{deluxetable*}

We have performed a detailed convergence study of our numerical models, the results of which will be presented in Papers II through IV of this series.  Here, we simply provide a brief summary of these tests. For all black hole accretion models, evaluated at the highest resolution (where applicable), the MRI quality factors in the vertical and azimuthal directions within the disk region are $Q_z \gtrsim 15$ and $Q_{\phi} \gtrsim 60$, respectively. These values are averaged within $50r_g$, where inflow equilibrium is approximately achieved, and are well within the criteria given in \citet{Hawley2011, Hawley2013}.  In addition, we have checked that the thermal scale height is resolved by at least 20 grid cells in the vertical direction.  We have also performed a resolution study for each of the single-loop simulations, along with one double-loop simulation (E07-a3-DL) that results in a thin disk structure.  We find the Maxwell stress, accretion rates, and structure of the disk are broadly consistent with the results presented here when the resolution is reduced by a factor of two in each dimension.

\subsection{Parameter Values}
\label{sec:parameter_values}

In this paper, we present the results from 10 simulations to study accretion onto stellar-mass black holes from sub- to super-Eddington regimes.  Our models span four different accretion rates (roughly $\sim 0.1\times$, $\sim 1\times$, $\sim 10\times$, and $\sim 100\times$ Eddington), two black hole spins (0.3 and 0.9375), and two magnetic field topologies (either single or double loops, with the former having a net poloidal flux at the midplane: see \autoref{fig:ini}). In subsequent papers, we will present a more comprehensive parameter survey.

Since the gas-radiation coupling terms are temperature-dependent, our simulations are no longer scale-free.  The models presented here correspond to models where the central black hole mass is set to 10 solar masses ($M=10M_{\odot}$). The initial density in the torus (and the resulting mass accretion rate) can be adjusted by specifying the density unit $\rho_0$: \autoref{tab:sim_result} lists the values used in each of our models.  One simulation (labeled E01-a3) is in the sub-Eddington regime, four simulations (labeled E08-a3, E08-a9, E07-a3-DL, and E09-a3-DL) are in the near-Eddington regime, and the remaining five (labeled E150-a9, E88-a3, E31-a3-DL, E15-a9, and E9-a3) correspond to super-Eddington accretion. Here, the naming convention for each simulation denotes the accretion rate (`E'), black hole spin (`a'), and initial magnetic field configuration (single-loop by default and `DL' for double-loop). All 10 simulations are run to at least $t=60000r_g/c$, while the sub-Eddington run and two near-Eddington runs with the double-loop magnetic configuration extend to $t=70000r_g/c$.

\section{Results} 
\label{sec:results}

\subsection{Time Evolution}

\begin{figure*}
    \centering
    \includegraphics[width=\textwidth]{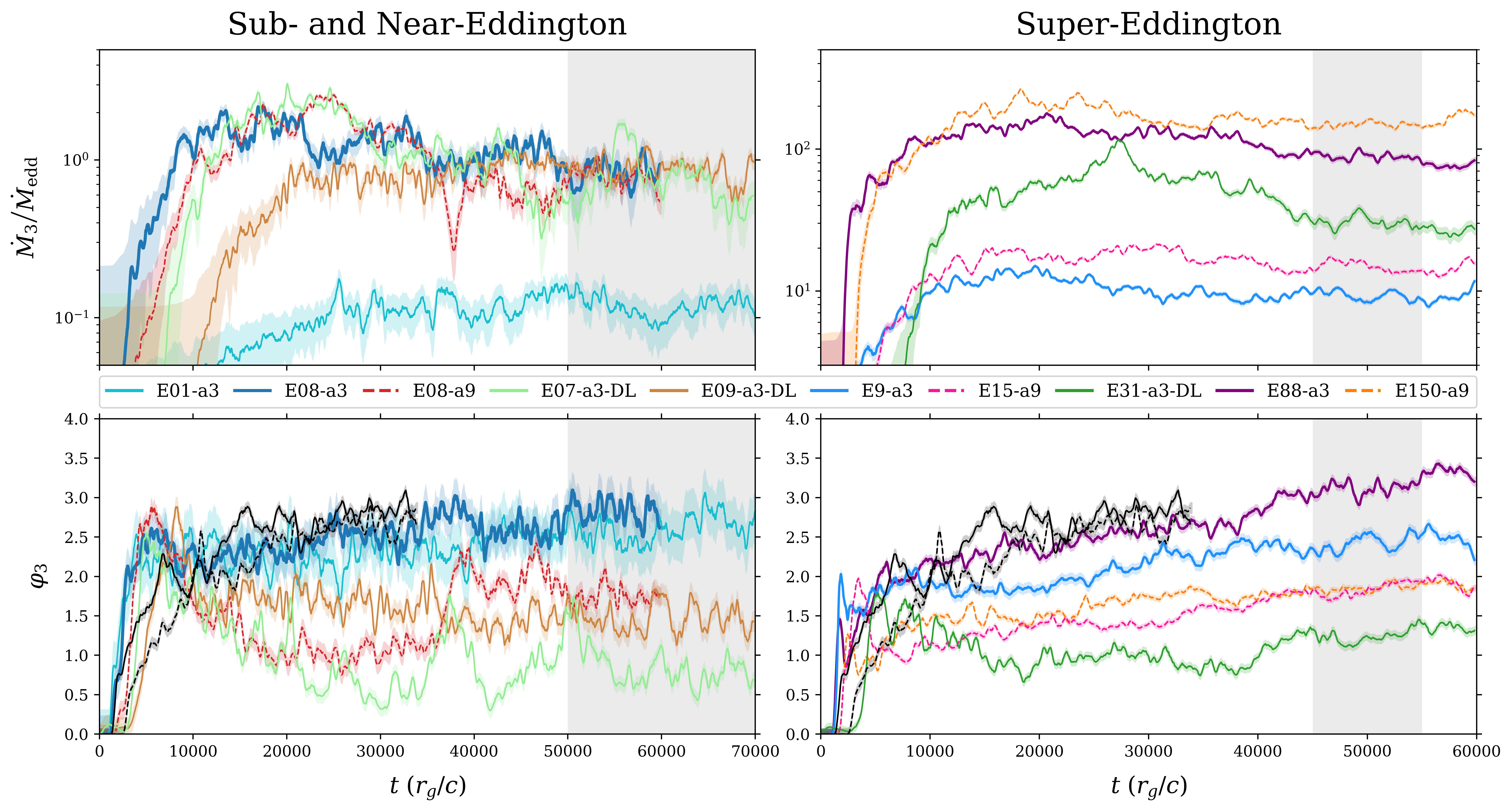}
    \caption{
    Time evolution of mass accretion rate and normalized magnetic flux at $r=3r_g$ (see equation~\ref{eq:acc_rate_and_mag_flux}).  In the lower panels, the black lines represent comparison simulations without radiation, with solid lines indicating low spin ($a=0.3$) and dashed lines indicating high spin ($a=0.9375$).  The gray-shaded regions indicate the time windows used for time-averaged analysis of the steady-state structure. 
    }
    \label{fig:hst_compare} 
\end{figure*}

In all of our models, development of the magnetorotational instability (MRI) leads to turbulence and angular momentum transport, ultimately establishing an accretion flow. \autoref{fig:hst_compare} shows the history of the mass accretion rate and magnetic flux near the horizon integrated over a spherical shell at $r=3r_g$, that is: 
\begin{subequations}
\begin{align}
    \dot{M}_{3} &= -\oint\rho u^r \sqrt{-g} d\theta d\phi
    \ ,
    \\
    \varphi_{3} &= \frac{1}{2\dot{M}^{1/2}}\oint |B^r| \sqrt{-g} d\theta d\phi
    \ ,
    \label{eq:mag_flux}
\end{align} 
\label{eq:acc_rate_and_mag_flux} 
\end{subequations}
where the subscript indicates that the integration is performed at $r=3r_g$.  Note that our simulations use Heaviside-Lorentz units (i.e., a factor of $4\pi$ is absorbed into the magnetic field definition), which in practice means the MAD state \citep{Tchekhovskoy2011} occurs for $\varphi_3 \approx 15$.  The mass accretion rate is normalized by the Eddington rate assuming a radiation efficiency of 0.1, that is $ \dot{M}_{\mathrm{Edd}}=4\pi G M/(0.1\kappa_Tc)$.  Each curve is averaged using a moving boxcar with a time window of $\Delta t=400r_g/c$. The shaded region along each curve represents the 1$\sigma$ variation within the local time window. 

The left two panels of \autoref{fig:hst_compare} show the evolution of models that achieve sub- and near-Eddington accretion rates.  Most cases show large fluctuations in the accretion rate $\dot{M}_{3}$ before it settles into a nearly steady state at $t > 40000r_g/c$.  In contrast, the magnetic flux near the horizon $\varphi_{3}$ saturates to its final value more quickly, although large fluctuations around that value continue throughout the evolution.  

The right two panels of \autoref{fig:hst_compare} show the evolution of super-Eddington models.  Final accretion rates that vary from roughly 10 to over 100  $\dot{M}_{\mathrm{Edd}}$ are achieved.  Both the mass accretion rate and flux near the horizon quickly settle into nearly steady values.  It is clear that super-Eddington accretion is much less variable than near-Eddington accretion, a point which we explore further below.

The gray-shaded region in each panel indicates the time window used for a detailed analysis of the steady-state structure.  In particular, the fourth and fifth columns of table \autoref{tab:sim_result} lists both the time-averaged mass accretion rate and flux near the horizon during this window for all of the models.  The near-Eddington models are all between $0.7-0.9 \dot{M}_{\mathrm{Edd}}$, while the super-Eddington models have a maximum accretion rate of almost $150 \dot{M}_{\mathrm{Edd}}$.  All simulations stay in the SANE regime with $\varphi_3 \leq 3.5$.  Those configured with a double-loop magnetic field (i.e., no net poloidal field at the equator) have a lower level of magnetic flux with $\varphi_3<2$.

It is notable that models configured with the single-loop magnetic field do not reach the MAD state, whereas in geometrically thick disks \citep{FM1976} this is always the case.   We find (Wong et al., in preparation) that even non-radiative models that begin with this field configuration in geometrically thin disks also do not go MAD, indicating this outcome is not caused by some aspect of the dynamics of radiation-dominated disks.  A variety of other authors have noted the emergence of the MAD state in thin disks; however, these models either begin with a geometrically thick (i.e. Fishbone-Moncrief) torus that is subsequently cooled \citep{Liska2022, Scepi2024, Dhang2025}, or they introduce additional vertical flux near the horizon, either in addition to or as a replacement for the single-loop field \citep{Avara2016, Morales-Teixeira2018, Curd2023}.  Our results are consistent with these results, namely without additional vertical flux, geometrically thin disks do not become MAD. 

\begin{figure*}
    \centering
    \includegraphics[width=\textwidth]{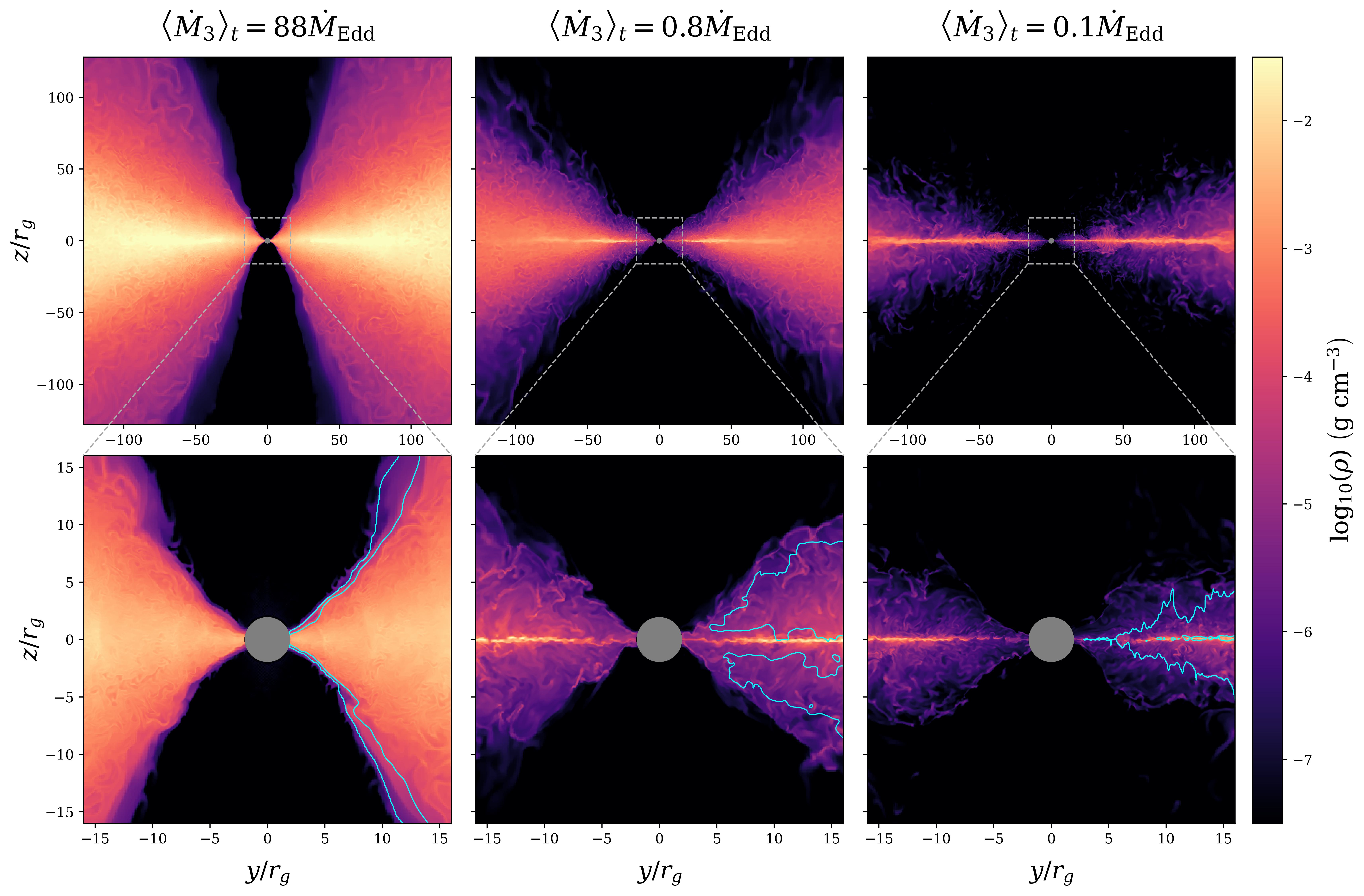}
    \caption{
    Poloidal slices of the mass density in radiation-dominated disks with the same black hole spin but different mass accretion rates.  The left column corresponds to a model with $\dot{M}_3 = 88 \dot{M}_{\mathrm{Edd}}$ (E88-a3), the middle column $\dot{M}_3 = 0.8 \dot{M}_{\mathrm{Edd}}$ (E08-a3), and the right column $\dot{M}_3 = 0.1 \dot{M}_{\mathrm{Edd}}$ (E01-a3).  The bottom row is a zoom-in to the inner regions.  Each model is shown at $t=60000r_g/c$ once a steady state has been reached.  The blue contour marks regions within the photosphere where the magnetic pressure is twice the thermal pressure, indicating magnetically dominated zones.
    }
    \label{fig:disk_Mdot}
\end{figure*}

From a more detailed analysis of the steady-state (given in Papers II through IV), we find inflow equilibrium is established roughly within $20$ to $60r_g$ in each case.  We show below our models have been run for at least one thermal time within this region. Most models generate more than 45\% of the total radiation luminosity within the region in inflow equilibrium, with the rest coming from larger radii in the disk.  The disks in all cases are radiation pressure dominated, with $\bar{P}_{r}/P_g$ varying from a few in the outer regions of models with the lowest accretion rates, to more than three orders of magnitude in the inner regions of highly super-Eddington models.  In the following subsections, we present further analysis of the steady state structure of these disks.

\subsection{Late-time Steady State}

\autoref{fig:disk_Mdot} plots poloidal slices of the mass density once steady state is achieved in three representative models initialized with a single-loop magnetic field: a super-Eddington disk with $\dot{M}_3 = 88 \dot{M}_{\mathrm{Edd}}$, a near-Eddington disk with $\dot{M}_3 = 0.8 \dot{M}_{\mathrm{Edd}}$, and a sub-Eddington disk with $\dot{M}_3 = 0.1 \dot{M}_{\mathrm{Edd}}$.  In each case a snapshot over large scales $\pm 128r_g$ is shown in the top row, and a zoom-in to the inner regions $\pm 16r_g$ is shown in the bottom row.  

Several important properties of these radiation-dominated flows are immediately obvious.  Firstly, the highly super-Eddington disk remains geometrically thick. In fact we find it has a structure which is strikingly similar to a non-radiating model (not shown, but presented in Paper II) that is run with exactly the same initial conditions and grid, but without solving the radiation transport equation, the major difference being the low-density funnel region in the super-Eddington disk is much wider. Secondly, both the near-Eddington and sub-Eddington cases result in a much thinner thermally supported layer overlaid by a magnetically dominated corona.   Finally, close inspection of the density snapshots in the inner regions reveal the density fluctuations in MRI turbulence in radiation-dominated flows are very large, especially compared to models without radiation, a result seen previously in local shearing box models \citep{Turner2003, JSD-RadTurb2013}.

To further investigate the structure of these flows, we identify regions corresponding to the bound accretion disk, unbound and outflowing wind, and collimated jet as follows.
The accretion disk is identified as the region with negative Bernoulli parameter, defined as
\begin{subequations}    
\begin{align}
    \mathrm{Be} &= -\frac{w_s u_t}{\rho} - 1
    \ , 
    \\
    w_s &= \begin{cases}
    \rho + \frac{\gamma}{\gamma-1}P_g + b^{\lambda}b_{\lambda}
    \qquad\quad\quad(\text{if } \tau_s \le 1)
    \\
    \rho + \frac{\gamma}{\gamma-1}P_g + b^{\lambda}b_{\lambda} + \frac{4}{3}\bar{E}_r
    \quad(\text{if } \tau_s > 1)\ ,
    \end{cases}
\end{align}
\end{subequations}
where $\bar{E}_r$ represents the radiation energy density in the fluid frame, and $\tau_s$ is the scattering optical depth vertically integrated from the edge of the simulation domain to the midplane.  Note that the Bernoulli parameter is modified by including radiative inertia in the optically thick region (i.e. $\tau_s>1$), where gas and radiation are well coupled.  In steady state, gas with a positive Bernoulli parameter is no longer gravitationally bound to the black hole, and therefore regions outside the zero contour we define as the outflowing wind. 

Even though none of our models enter the MAD state, we nonetheless find powerful relativistic jets are formed in some cases, especially when the central black hole has large spin and accumulates vertical magnetic field.  In regions where the Bernoulli number is positive, we identify the jet region by tracing the velocity streamlines in the time-averaged flow.  The jet consists of those streamlines which originate from the strongly magnetized inner region, where the magnetic energy exceeds the rest-mass energy, and extend to the edge of the simulation domain.  The outermost streamline of this region is used to define the boundary between the jet and wind. 

\begin{figure*}
    \centering
    \includegraphics[width=\textwidth]{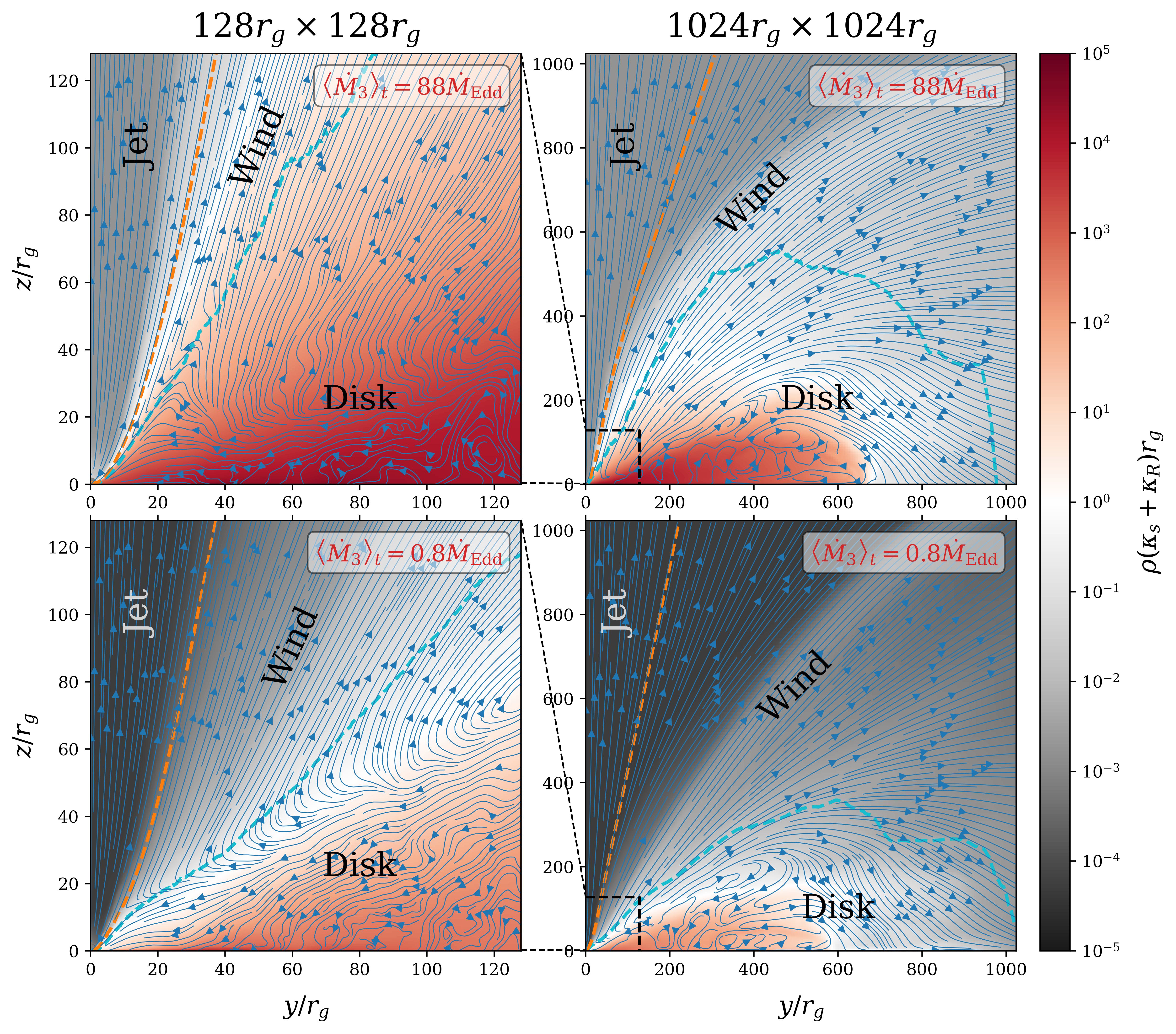}
    \caption{Disk, wind, and jet regions that emerge in our radiation-dominated disk models.  The upper panels illustrate the super-Eddington case $\dot{M}_3 = 88 \dot{M}_{\mathrm{Edd}}$ (E88-a3), while the lower panels show a near-Eddington case $\dot{M}_3 = 0.8 \dot{M}_{\mathrm{Edd}}$ (E08-a3).  Colors show the local flux-mean optical depth per gravitational radius.  The blue streamlines show the gas velocity.  The cyan dashed lines mark the contour of zero Bernoulli parameter: regions inside this contour are gravitationally bound.  The orange dashed lines outline the jet zone, defined by the outermost velocity streamline that originates from the strongly magnetized inner region where magnetic energy equals rest mass energy.     
    }
    \label{fig:partition}
\end{figure*}

\autoref{fig:partition} shows the result of partitioning two of our simulations according to the criteria described above, using the models with $\dot{M}_3 = 88 \dot{M}_{\mathrm{Edd}}$ and $\dot{M}_3 = 0.8 \dot{M}_{\mathrm{Edd}}$ shown in \autoref{fig:disk_Mdot}.  The left panels show a zoom-in to the inner regions of the flow, while the right panels show the entire simulation domain.
The orange dashed lines in each panel delineate the jet boundaries, while the cyan dashed lines mark the contour of zero Bernoulli number, and therefore the region of bound gas we identify as the accretion disk.  Blue lines show the velocity streamlines, clearly showing outward flow in the jet and wind regions and more chaotic and turbulent flow within the disk.  Note that the profiles are time-averaged, and the color scale in this figure shows the flux-mean optical depth per gravitational radius $r_g$ and not the gas density.  Red regions are very optically thick, while gray to black regions are optically thin.  The photosphere will be located somewhere near the boundary of light gray and white regions.  Note there are significant differences in the optical depth of the outflows generated in these two case, with the super-Eddington outflow being mostly optically thick, while the near-Eddington case is more nearly optically thin.

We also note that even in the disk region composed of bound gas, there is a coherent outflow from small to large radii.  Thus mass is lost from the inner accretion flow, and is recycled into the disk at large radii where it may be accreted at late times.  Such recycling flows have also been observed in gas pressure dominated disks \citep{Jacquemin-Ide2021}.

In the following subsections, we analyze the properties of the disk and outflow regions in more detail.

\subsection{Disk properties}
\label{sec:disk_properties}

\subsubsection{Vertical structure}
\label{sec:vert_struct}

To investigate the vertical structure of the disk in each case, we measure the scale height of both the density and magnetic pressure measured at $r=10r_g$ defined by 
\begin{subequations}    
\begin{align}
    H &= \frac{\displaystyle\int_{\mathrm{disk}} |z|\left<\rho\right>_{\phi,t} dz}{\displaystyle\int_{\mathrm{disk}} \left<\rho\right>_{\phi,t} dz}
    \ ,
    \\
    H^{(\mathrm{mag})} &= \frac{\displaystyle\int_{\mathrm{disk}} |z|\left<b^{\lambda}b_{\lambda}\right>_{\phi,t} dz}{\displaystyle\int_{\mathrm{disk}} \left<b^{\lambda}b_{\lambda}\right>_{\phi,t} dz}
    \ ,
\end{align}
\end{subequations}
where $\langle \rangle_{\phi, t}$ denotes azimuthal and time averages.
In columns six and seven of \autoref{tab:sim_result}, we report the density and magnetic pressure scale height for each model.  The former shows a clear trend with Eddington ratio, with the disk thickness decreasing from $H/r \approx 0.25$ at $150 \dot{M}_{\mathrm{Edd}}$ to 0.02 at $0.1 \dot{M}_{\mathrm{Edd}}$. In contrast, the magnetic pressure scale height varies by much less, only by a factor of about $2\times$ over the same range.  In every case the magnetic pressure scale height is larger than that of the density.  At low mass accretion rates, it can be $10\times$ larger.  This results in the formation of a strongly magnetized corona above the dense disk midplane.  At low accretion rates, the corona is above the photosphere, and this results in much higher gas temperatures.  This important result has been studied previously \citep{JSD2014} in non-relativistic models, and therefore is a basic property of MRI-driven accretion flows.

Note also that the initial magnetic field geometry affects the late-time vertical structure of the disk.  The near-Eddington model that begins with the double-loop geometry (E09-a3-DL) has a density scale roughly $5\times$ larger than an equivalent model that beings with a single loop (E08-a3).  The effect of the initial magnetic field is explored further in \autoref{sec:disk_struct}.

The radial profile of disk temperature is computed as a vertical average of the time- and azimuthally averaged pressure and density within the disk,
\begin{equation}
    T_g^{\mathrm{(disk)}} = \frac{\mu m_p}{k_B} \frac{\displaystyle\int_{\mathrm{disk}} \left<P_g\right>_{\phi,t} dz}{\displaystyle\int_{\mathrm{disk}} \left<\rho\right>_{\phi,t} dz}
    \ .
\end{equation}
We find the radial profile of the temperature in the inner disk follows a power law in all cases.  Therefore, in column eight of \autoref{tab:sim_result} we list the amplitude and best fit power-law index for the temperature (denoted with a tilde) in each simulation within the inflow equilibrium region.  In general, the disk temperature increases uniformly with higher accretion rates, following a shallow power-law index close to 0.25. As expected for accretion onto stellar-mass black holes, the midplane temperatures typically exceed $10^7$~K.

The outgoing luminosity generated by the disk is calculated by integration of the outgoing flux over a cylinder outside the scattering photosphere , 
\begin{equation}
    L = -\oint_{\mathrm{ph}} \left<R^z_{\ t}ds + R^s_{\ t}dz\right>_{t} sd\phi
    \ ,
    \label{eq:luminosity}
\end{equation}
where $s=\sqrt{x^2 + y^2}$ is the cylindrical radius.  The height of the cylinder is defined by the photosphere at each corresponding radius.  The radiation efficiency at any given radius is then
\begin{equation}
    \eta^{(\mathrm{rad})} = L/(\dot{M}c^2)
    \ .
    \label{eq:radiative_eff}
\end{equation}
Note that $\dot{M}$ is meaningful only in the region of inflow equilibrium.  For regions beyond this radius, we adopt the same $\dot{M}$ for the evaluation.  

The radiation efficiency $\eta^{(\mathrm{rad})}_{20}$ measured at a radius of $20r_g$ is given in the ninth column of \autoref{tab:sim_result}.  We find it is strongly inversely correlated with the mass accretion rate, dropping from over 5\% at $\dot{M} = 0.1 \dot{M}_{\mathrm{Edd}}$ to less than 0.5\% at $\dot{M} = 150 \dot{M}_{\mathrm{Edd}}$.  Models with the initial double-loop magnetic field configuration appear to have a slightly lower radiative efficiency at the same mass accretion rate as models that start with a single loop.

\begin{figure*}
    \centering
    \includegraphics[width=0.96\textwidth]{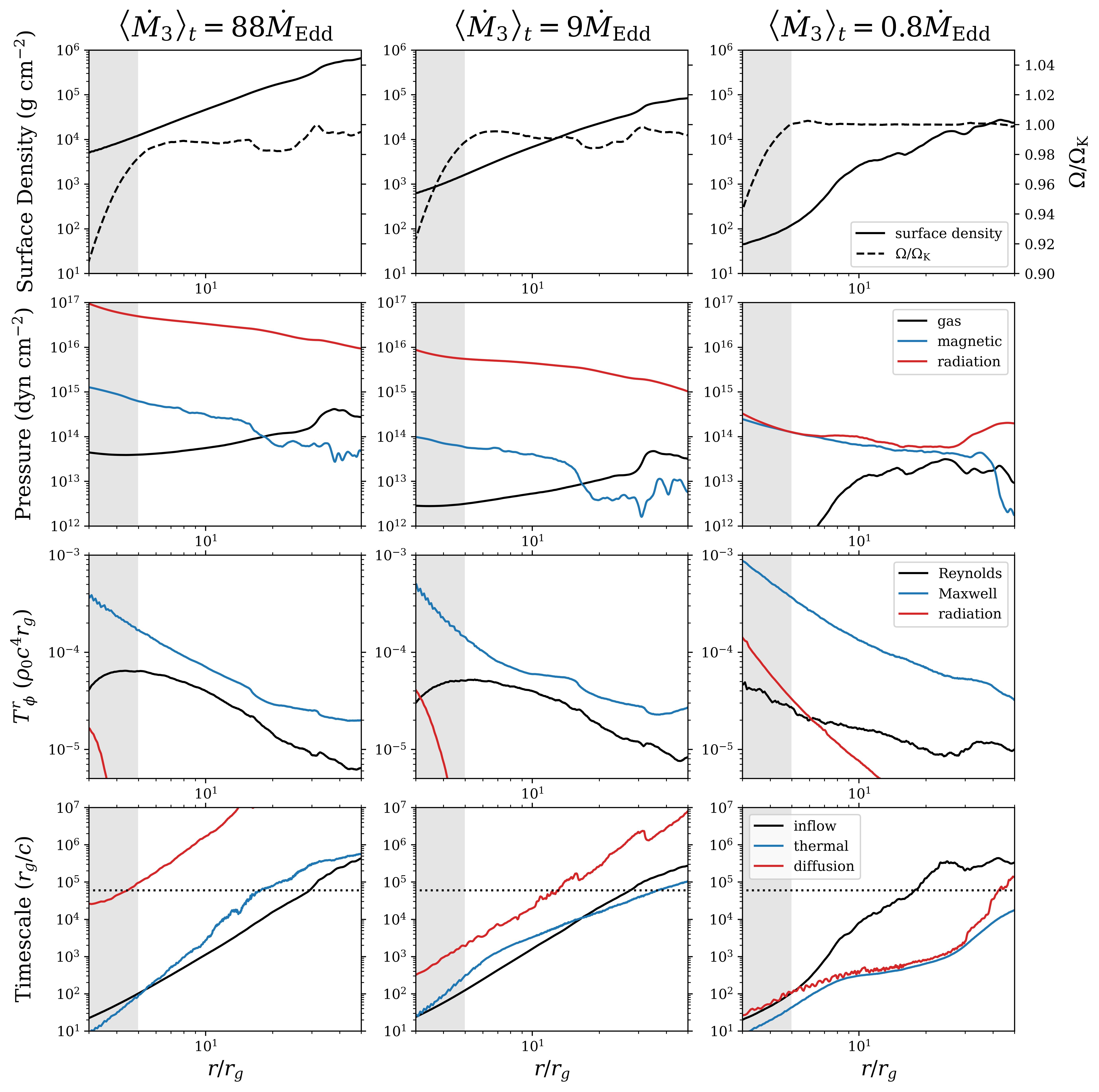}
    \caption{Radial profiles of (from top to bottom) surface density and midplane angular frequency normalized by the Keplerian value, midplane pressures, disk-averaged stresses, and various timescales. Gray shading denotes the plunging region inside the ISCO.  Each column shows the results from a different accretion regime (models E88-a3, E9-a3, and E08-a3).  These models all have the same spin ($a=0.3$) and are initialized with a single-loop magnetic field configuration. 
    }
    \label{fig:hori1d_compare}
\end{figure*}

We discuss the radiative efficiency of our models further in \autoref{sec:EffVar}, especially with regards to the implications for the observed luminosity of super-Eddington accretion.  We note that a very wide range of radiation efficiencies have been reported in the literature, with some authors reporting efficiencies of near 100\% for both near-Eddington \citep{Fragile2023, Curd2023} and super-Eddington \citep{McKinney2017} flows, albeit for the MAD state.  More moderate efficiencies in the range of 30--70\% have been reported by \citet{Fragile2025}.  At the same time, efficiencies of $\sim10$\% have also been reported \citep{McKinney2014, McKinney2015, Sadowski2016, Morales-Teixeira2018}.  These latter values are much closer to those we find here.  Further discussion of the comparison of our results to previous models is given in \autoref{sec:previous-work}.

\subsubsection{Radial profiles}

In \autoref{fig:hori1d_compare}, we present the radial profiles of time-averaged quantities for three models (E88-a3, E9-a3, and E08-a3) corresponding to mass accretion rates of $88 \dot{M}_{\mathrm{Edd}}$, $9 \dot{M}_{\mathrm{Edd}}$, and $0.8 \dot{M}_{\mathrm{Edd}}$ respectively, but with the same spin.  The plunging region inside the ISCO is denoted with gray shading in each panel.  The top row shows the surface density (solid lines) and midplane angular frequency (dashed lines) normalized by the orbital frequency
\begin{equation}
    \Omega_{\mathrm{K}}=\left[\left(\frac{r}{M}\right)^{\frac{3}{2}}+a\right]^{-1}
    \ .
\end{equation}

The surface density profiles increase roughly as $r^{3/2}$ in each case.  This seems to be steeper than expected for slim-disk models of super-Eddington accretion, e.g., \citet{Sadowski_2011}, especially at the lower accretion rates.  Our measured profiles are quite similar to those found previously for super-Eddington accretion onto both stellar-mass \citep{Jiang2014b} and supermassive \citep{Jiang2019ApJ880} black holes. 
The profiles of angular frequency are near-Keplerian except close to the ISCO in each case, indicating that pressure support is not important in the radial direction.

The second row compares the midplane gas, magnetic, and radiation pressures.  Here, the magnetic pressure is calculated as $b^{\lambda}b_{\lambda}/2$, while the radiation pressure is given by $\bar{E}_r/3$ in the fluid frame.  Radiation pressure dominates across all accretion regimes, particularly in the highly super-Eddington case where it is more than three orders of magnitude larger than the gas pressure close to the black hole.  Magnetic pressure is larger than gas pressure inside of $r \lesssim 20r_g$ in the super-Eddington models. In the near-Eddington regime, magnetic pressure becomes comparable to the radiation pressure. 

The third row of \autoref{fig:hori1d_compare} demonstrates how angular momentum is transported in the radial direction by plotting the disk-averaged stresses defined as follows: 
\begin{subequations}    
\begin{align}
    \left(T_{\mathrm{Rey}}^{\mathrm{turb}}\right)^{r}_{\ \phi} &= w_{\mathrm{disk}} u^r u_{\phi} - \frac{\left<w_{\mathrm{disk}}u^r\right>_{\phi} \left<w_{\mathrm{disk}}u_{\phi}\right>_{\phi}}{\left<w_{\mathrm{disk}}\right>_{\phi}}  
    \ ,
    \\
    \left(T_{\mathrm{Max}}\right)^{r}_{\ \phi} &= -b^r b_{\phi}
    \ ,
    \\
    \left(R_{\mathrm{diff}}\right)^{r}_{\ \phi} &= R^{r}_{\ \phi} - \frac{4}{3}\bar{E}_r u^r u_{\phi} g
    \ ,
\end{align}
\end{subequations}
where $w_{\mathrm{disk}} = \rho + \gamma P_g/(\gamma-1) + b^{\lambda}b_{\lambda} + 4\bar{E}_r/3$ represents the total enthalpy.  The Maxwell stress $\left(T_{\mathrm{Max}}\right)^{r}_{\ \phi}$ (blue lines) dominates the outward radial angular momentum transport across all radii and accretion rates.  The turbulent Reynolds stress $\left(T_{\mathrm{Rey}}^{\mathrm{turb}}\right)^{r}_{\ \phi}$ (black lines) is significant, though sub-dominant, in the super-Eddington regime but contributes minimally in the near-Eddington accretion.  The radiation stress $\left(R_{\mathrm{disk}}\right)^{r}_{\ \phi}$ (red lines) consistently remains a minor component in angular momentum transport across all regimes.  These stresses are measured as averages over the disk, and therefore are weighted to the optically thick regions.  Radiation viscosity may be larger in the optically thin corona, especially for sub-Eddington accretion \citep{Jiang2019ApJ885}.

The profiles of the stress within the ISCO (gray shaded regions) are of interest.  In particular, there is no edge to the Maxwell stress at the ISCO, instead it increases monotonically inwards. On the other hand, the Reynolds stress does decline significantly inside the ISCO in some cases, presumably due to the radial stretching of turbulent eddies. The fact the total stress does not decline to zero at or near the ISCO is in contrast to the typical assumptions made in steady-state $\alpha-$disk models \citep{Novikov1973}, and has been noted before as a robust property of turbulent GRMHD models \citep[e.g.][]{KrolikHawley2002} in which there is nothing special about the ISCO nor the sonic point calculated from the time- and azimuthally averaged flow.  Recent analytic models of the flow inside the ISCO \citep{Mummery2024} seem to fit the average radial profiles in non-radiative models quite well \citep{MummeryStone2024}.  More detailed comparisons with these radiation-dominated solutions will be given in future papers in this series.

The last row of \autoref{fig:hori1d_compare} compares the inflow, thermal, and diffusion timescales in order to investigate energy transport in each model. The thermal time measures the cooling time of the disk including both radiation diffusion and advection normal to the emitting surface, while the diffusion time measures the same quantity but only including diffusion.  The precise formulae used to calculate these timescales are given in \autoref{appendix:timescales}. Horizontal dotted lines in each panel show the total simulation time, regions with inflow times that exceed this may not be relaxed. As shown previously \citep{JSD-ThermalStab2013, Jiang2014b}, turbulent transport of radiation energy can greatly exceed diffusion, and therefore determine the overall cooling time.  We reproduce this result here, with the thermal time (blue line) being much smaller than the diffusion time (red line) at all radii in the super-Eddington cases. 

\begin{figure*}
    \centering
    \includegraphics[width=\textwidth]{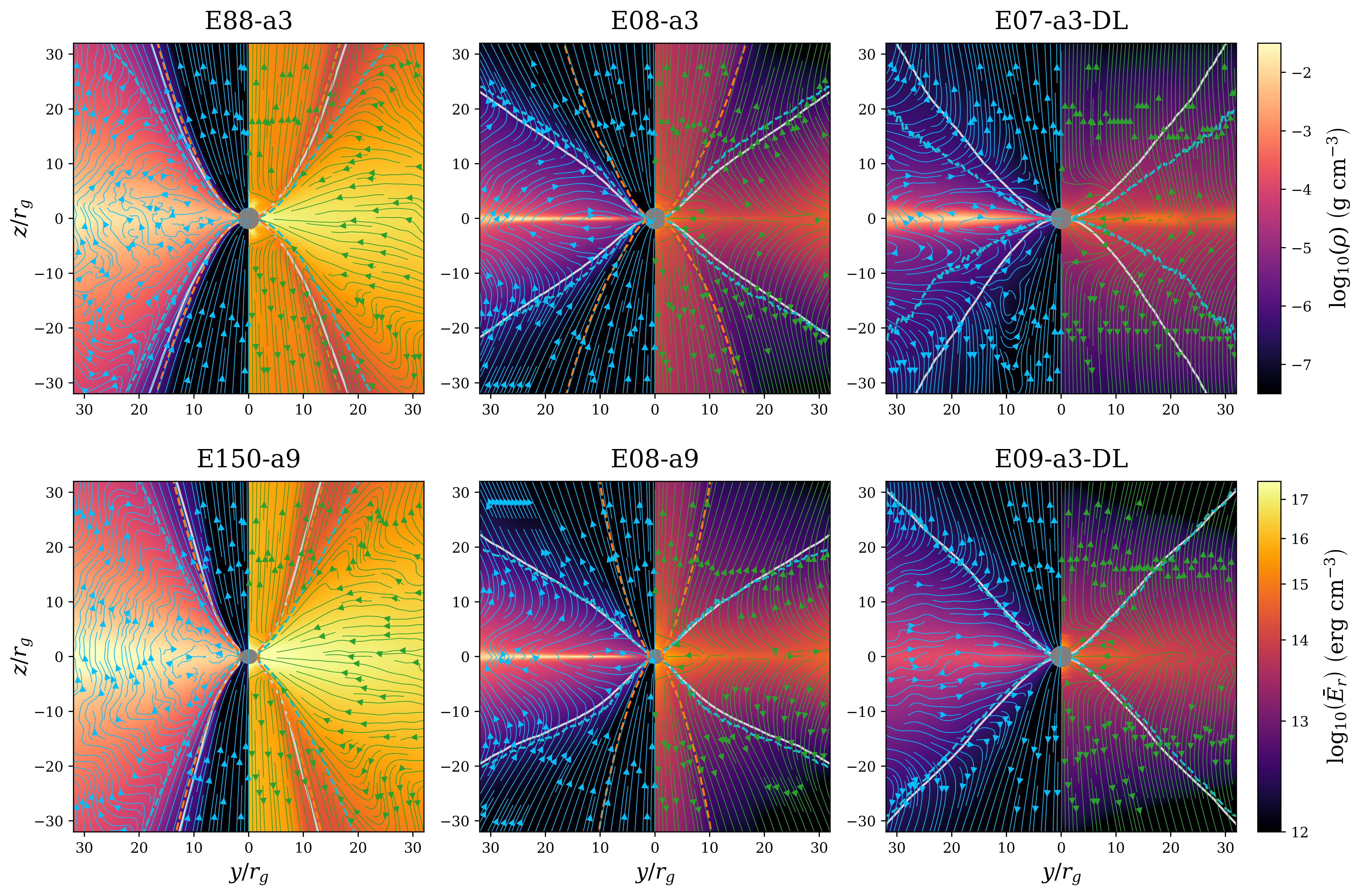}
    \caption{Temporal and azimuthal averages of gas density (left half of each panel) and fluid-frame radiation energy density (right half of each panel) for six different models, spanning different accretion rates and black hole spins.  Blue streamlines in the left panels trace the magnetic field, while green streamlines in the right panels indicate coordinate-frame radiation flux.  The disk body is enclosed by blue dashed lines, marking the contour of zero Bernoulli parameter.  Grey solid lines denote the scattering photosphere, and orange dashed lines outline the jet region.  Note that no orange dashed lines appear in the two models in the right column (i.e. E07-a3-DL and E09-a3-DL) as these do not form a steady jet.     
    }
    \label{fig:profile2d}
\end{figure*}

Since photons can be transported by both diffusion and advection, the photon trapping radius $r_{\mathrm{tr}}$ \citep{Begelman1978} is more appropriately defined as the point where the inflow time becomes less than the thermal time.  Balancing the radial inflow and photon diffusion times predicts $r_{\mathrm{tr}}/r_g \sim (\dot{M}/\dot{M}_{\mathrm{Edd}})(H/r)$ \citep[e.g.][]{Ohsuga2002}, where $H$ is the density scale height.  Applying this expectation to our results is problematic since as shown in \autoref{fig:hori1d_compare} the transport of radiation is dominated by advection in these turbulent accretion flows rather than diffusion.  Nonetheless, from the figure we note the trapping radius increases from about $20r_g$ to more than $50r_g$ as the accretion rate increases from $9$ to $88\dot{M}_{\mathrm{Edd}}$.  Thus there is a clear trend that the trapping radius increases significantly at higher mass accretion rates, but the precise dependence requires a much finer sampling of models at different accretion rates.

\subsection{Jet and outflow properties}
\label{sec:wind_jet_properties}

To further investigate the properties of outflows and relativistic jets formed in the strongly magnetized funnel region, in \autoref{fig:profile2d} we plot time- and azimuthal averages of the mass (left) and radiation energy density (right), along with streamlines of the magnetic field (blue) and radiation flux (green), for six models that span mass accretion rate from $0.7 \dot{M}_{\mathrm{Edd}}$ to $150 \dot{M}_{\mathrm{Edd}}$, as well as different spins and initial magnetic field geometries.  Also shown as a cyan dashed line in each panel are the zero contour of the Bernoulli parameter (which marks the boundary between bound and outflowing gas), as well as the boundary of the jet region (when present) as an orange dashed line.

In addition, in order to further investigate the energetics of the wind and jet, we measure the kinetic, internal, and magnetic energy carried in the outflows as
\begin{subequations}    
\begin{align}
    \dot{E}_{\mathrm{kin}}^{\mathrm{(zone)}} &= \oint_{\mathrm{zone}} \left<-\rho u^r\left(u_t+\sqrt{-g_{tt}}\right)\right>_{t} \sqrt{-g} d\theta d\phi    
    ,
    \\
    \dot{E}_{\mathrm{egas}}^{\mathrm{(zone)}} &= \frac{\gamma}{\gamma-1}\oint_{\mathrm{zone}} \left<-P_g u^r u_t \right>_{t} \sqrt{-g} d\theta d\phi    
    \ ,
    \\
    \dot{E}_{\mathrm{mag}}^{\mathrm{(zone)}}  &= \oint_{\mathrm{zone}} \left< -b^{\lambda}b_{\lambda}u^r u_t + b^r b_t \right>_{t} \sqrt{-g} d\theta d\phi    
    \ ,
\end{align}
\end{subequations}
where the superscript `zone' refers to either the wind or jet region.  In order to separate the energy carried by radiation versus the fluid, we do not include the contribution from radiation in these fluxes.  The energy carried by free-streaming radiation in the optically-thin funnel regions has already been discussed (see equation~\ref{eq:luminosity} in \autoref{sec:vert_struct}). The efficiency of wind or jet can then be defined as
\begin{equation}
    \eta^{(\mathrm{zone})} = \frac{\dot{E}_{\mathrm{kin}} + \dot{E}_{\mathrm{egas}} + \dot{E}_{\mathrm{mag}}}{\dot{M}c^2}
\end{equation}
at any given radius.  We use the same $\dot{M}$ as in the calculation of radiation efficiency (see equation~\ref{eq:radiative_eff}) for direct comparison.  Using these diagnostics, the time- and azimuthally averaged efficiency of the wind (measured at $200r_g$) and jet (measured at $20r_g$) are given in columns ten and eleven in  \autoref{tab:sim_result}.  We select these radii since the wind and jet power profiles roughly become constant beyond them.

Several features are evident in \autoref{fig:profile2d}.  The two models in the left column both have super-Eddington accretion rates but different spins. Both models produce powerful jets (see \autoref{tab:sim_result}), with the jet efficiency nearly twice as large with higher spin.  Outflows carry much more energy than radiation in both cases (more than $3\times$ in the high spin case).  The green streamlines in the right panels indicate radiation flux is radial (inwards) inside disk demonstrating that photons are trapped.  However, in the optically thin funnel region the radiation streamlines are nearly radially outwards.  The photosphere is close to the jet boundary, meaning that outflowing winds (between the jet boundary and Bernoulli zero contour) are optically thick.  The magnetic field is largely tangled except in funnel region where it shows a clear hour-glass shape.

The two models in the center column both have near-Eddington accretion rates and start from the single-loop initial magnetic field, but have different spins.  In this case the disk and jet structure are similar.  The higher spin model has a more collimated jet with a smaller opening angle.  From \autoref{tab:sim_result}, wind power is much larger in high-spin case, but jet power is about the same since the low spin case has more magnetic flux on the horizon.  The radiative efficiency of these models is much larger than the super-Eddington cases, so now radiation carries more energy than the jet and wind.  Inside the disk the radiation flux is nearly vertical, reflecting the small contribution of advection to the cooling rate (see discussion of \autoref{fig:hori1d_compare}).  The outflowing winds are now mostly optically thin.

The two panels in the right column both have near-Eddington accretion rates and the same spin, but start from the double-loop initial magnetic field.  These models do not form steady jets, as both have small flux on the horizon.  Of particular interest is the model in the lower right panel with $\dot{M} = 0.9 \dot{M}_{\mathrm{Edd}}$.  The density scale height for this model is much larger than the nearly identical model that starts from the single-loop configuration (E08-a3), showing that the initial magnetic field structure (and in particular whether there is a net vertical flux at the midplane) is important in determining the long-term steady state of the accretion flow.

\begin{figure}
    \centering
    \includegraphics[width=\columnwidth]{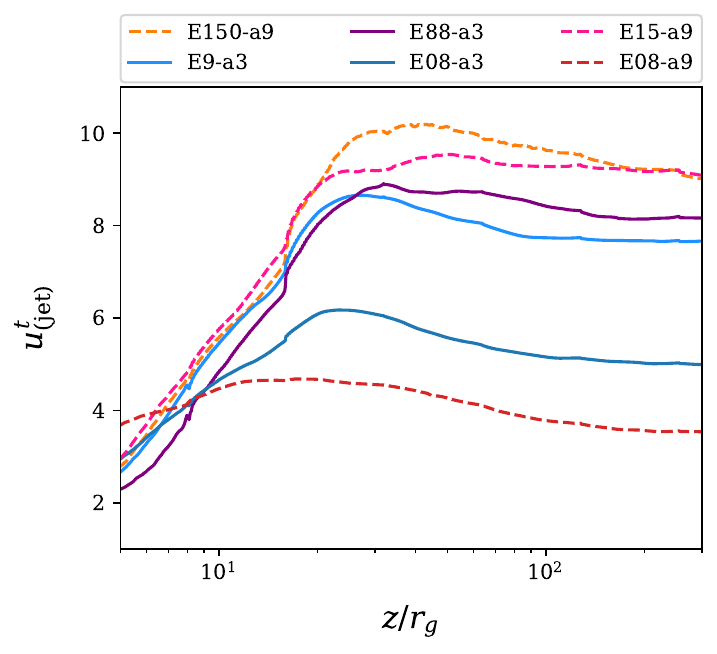}
    \caption{Time component of the jet four-velocity. The jet profile is temporally and azimuthally averaged. The four-velocity is spatially averaged using the density weighting along the jet spine.  
    }
    \label{fig:ut_jet}
\end{figure} 

\begin{figure*}
    \centering
    \includegraphics[width=\textwidth]{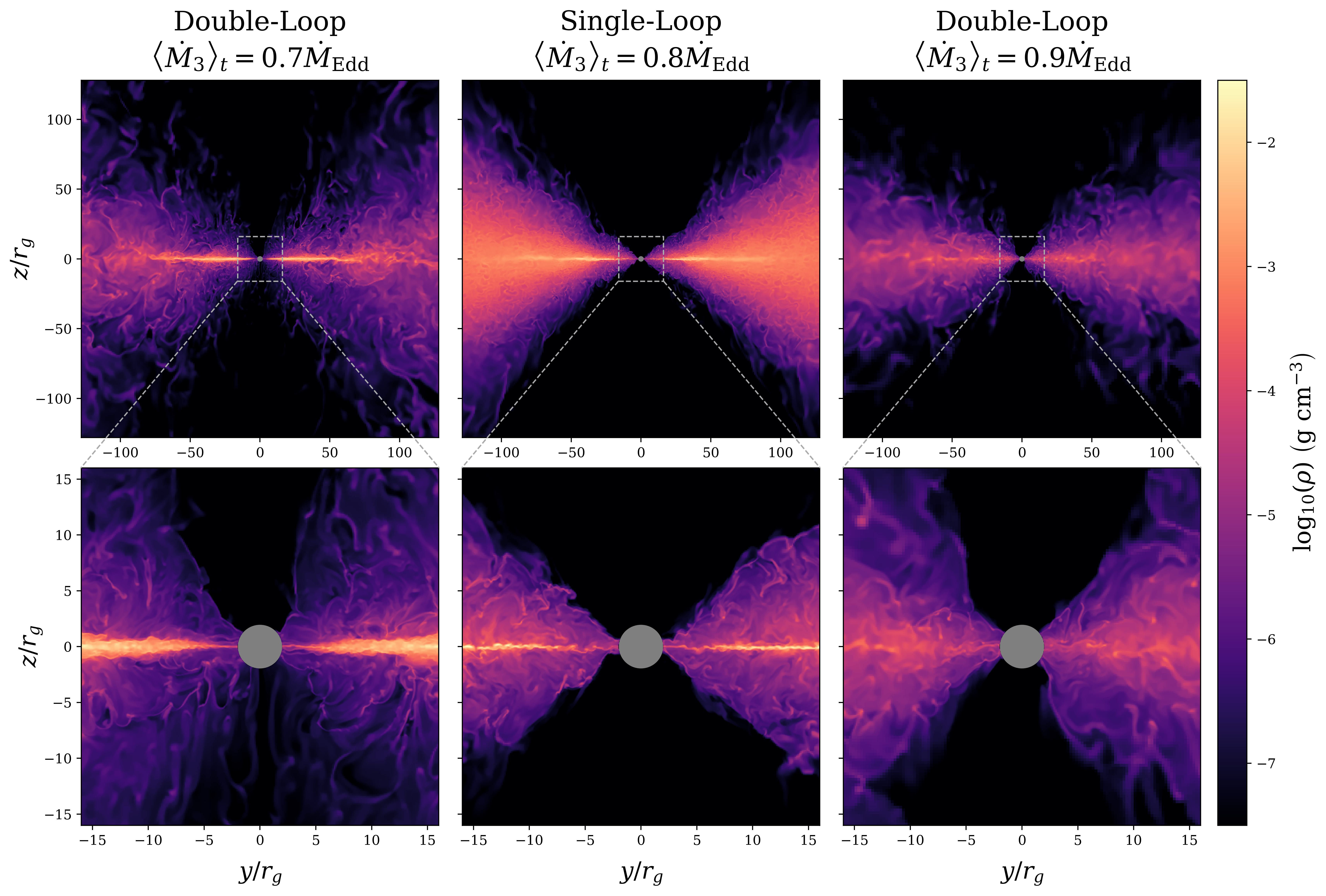}
    \caption{
    Poloidal slices of the mass density in radiation-dominated disks with the same spin and roughly the same mass accretion rate, but different initial magnetic field geometries.  The left and right columns show models which begin with a double-loop configuration (no vertical magnetic flux at the midplane initially), while the middle column shows a model with a single-loop configuration.  From left to right the models shown are E07-a3-DL, E08-a3 and E09-a3-DL. The bottom row is a zoom-in to the inner regions. Each model is shown at $t=60000r_g/c$.  
    }
    \label{fig:disk_Bfield}
\end{figure*}

To investigate the acceleration and terminal velocity of the jet, we compute the time component of the four-velocity along the jet spine, defined with an opening angle of $\theta_{\mathrm{sp}}=\mathrm{min}(\theta^{(\mathrm{jet})}, r_g/r)$, where $\theta^{(\mathrm{jet})}$ is the angle subtended by the jet region at the radius $r$.  Within the jet spine, the coordinate-frame Lorentz factor is calculated from the horizontally-averaged momentum
\begin{equation}
    u^t_{\mathrm{(jet)}} = \frac{\displaystyle\oint_{\mathrm{sp}} \left<\rho u^t\right> \sqrt{-g} d\theta d\phi}{\displaystyle\oint_{\mathrm{sp}} \left<\rho\right> \sqrt{-g} d\theta d\phi}
    \ ,
\end{equation}
where the integration is taken over the opening angle of the jet.  \autoref{fig:ut_jet} plots the profile of $u^t_{\mathrm{(jet)}}$ along the jet spine.  In each case, the jet accelerates uniformly to $20-30r_g$ and then slightly decelerates thereafter.  Analysis of the forces acting on the fluid (presented in Papers II through IV) shows that the jet is launched and accelerated to its peak velocity mostly by the outward magnetic tension force near the black hole, driven by the Blandford-Znajek process \citep{Blandford1977, Kinoshita2017}.  During this phase, the energy outflow is dominated by the Poynting flux.  Deceleration beyond the peak is also primarily magnetic due to the relaxation of helical field lines.  As the jet transitions away from being magnetically dominated, the radiation drag force takes over and further decelerates the flow.  The largest radiation drag occurs in the super-Eddington models.  Capturing the radiation-drag forces on the jet in the funnel region correctly is one of the strengths of the full-transport algorithms adopted in this work.  More detailed analysis will be presented in our subsequent papers. 

In column twelve of \autoref{tab:sim_result}, we report the maximum Lorentz factor along the jet spine, max$\left(u^t_{\mathrm{(jet)}}\right)$, for each model.  Only models that begin with the single-loop field geometry accumulate enough vertical flux at the horizon to generate a relativistic jet.  Models with a more rapidly spinning black hole tend to produce faster (more energetic) jets (as expected), and there is an indication that super-Eddington disks produce the most powerful jets.  This may be because the narrow funnel region provides more collimation and faster jets, despite the larger radiation drag in these more luminous flows. 

Column thirteen in \autoref{tab:sim_result} also lists the time- and azimuthally-averaged mass outflow rate at the outer edge of the domain $r=1024r_g$ measured in units of $\dot{M}_{3}$, the mass accretion rate at $r=3r_g$.  This ratio is remarkably constant at near unity across over three orders of magnitude in accretion rates.  Moreover, the near- and sub-Eddington models have the largest values, although of course the super-Eddington models have much larger mass loss rates when measured in absolute terms (e.g. $\dot{M}_{\mathrm{Edd}}$).  The lower luminosity models also have larger efficiencies $\eta_{200}^{\mathrm{(wind)}}$, perhaps reflecting the larger contribution of magnetically-driven outflows in these models.

\subsection{Dependence on initial field geometry}
\label{sec:disk_struct}

The steady-state structure of the accretion flow that emerges from these models depends on the initial magnetic field geometry (single versus double loops).  To illustrate this more clearly, in \autoref{fig:disk_Bfield} we plot poloidal slices of the density from three models with roughly the same (near-Eddington) accretion rate and black hole spin ($a=0.3$) but with different initial field geometries.  
The run with no poloidal flux at the midplane (right, starting from a double field loop initial condition) forms a geometrically thick, magnetic pressure supported disk.  In contrast, models that either start with or develop net vertical flux (center and left) produce a much thinner gas pressure supported disk with a magnetically supported envelope. 
When combined with the models of super-Eddington accretion disks, we conclude there are three types of accretion disks evident in our models, as discussed below. 

\subsubsection{Super-Eddington thick disks}
\label{sec:super_edd_disk}

Regardless of the initial field geometry, each of the five super-Eddington accretion models result in a geometrically thick, radiation pressure supported disk (e.g. left column in \autoref{fig:profile2d}; see also \citealt{JiangDai2024}).  The primary source of angular momentum transport is via Maxwell stress and turbulent Reynolds stress, with the former dominating. Radiation viscosity provides very little contribution.  Turbulence in the disk produces large density fluctuations, characteristic of radiation pressure dominated plasmas \citep{Turner2003, JSD-RadTurb2013}.

The radiative efficiency in this regime is very small, and in fact the outward energy flux associated with an optically thick wind and a relativistic jet (when the black hole spin is large) can exceed that due to radiation by factors of a few.  Even though none of our super-Eddington models reach the MAD regime, those that begin with a single-loop magnetic field geometry still produce powerful jets. The total mass loss in the wind is comparable to the mass accretion rate at the horizon.

Vertical transport of energy in the disk is dominated by turbulent advection rather than radiative diffusion.  Nonetheless the trapping radius for photons, where the inflow time is less than the thermal time, extends to tens of $r_g$.  There is no sign of thermal or viscous instability in the inner regions of the disk.

\subsubsection{Gas pressure supported thin disks}
\label{sec:gasp_disk}

At near-Eddington accretion rates, simulations that start with a single-loop field geometry form a very thin (e.g. $H/r \sim 0.02$) dense disk at the midplane, with an extended magnetically dominated ($H^{(\mathrm{mag})}/r \sim 0.25$) corona above.  In addition, one model beginning with a double-loop field geometry (E07-a3-DL) also forms a similar structure (see the center and left panels in \autoref{fig:disk_Bfield}).  This structure is very similar to the ``puffy disk'' observed in the radiation-dominated models of \citet{Lancov2019, Wielgus2022, Fragile2023}.  Note that model E07-a3-DL eventually develops net vertical magnetic flux near the horizon generated by the combined action of turbulence and disk outflows.

The vertical structure of the thin disk is mostly supported by gradients of the gas pressure, even though radiation pressure dominates by magnitude.  The magnetic corona is supported by a combination of both radiation and magnetic pressure.  Angular momentum transport is dominated by Maxwell and Reynolds stresses.  Because the thin gas layer at the midplane contains vertical magnetic flux, vertical transport via the $b^{z}b_{\phi}$ component of the stress tensor is also important.  This leads to most inward accretion occurring in the upper magnetized layers, with gas moving outwards at the midplane in some regions very reminiscent of the vertical structure observed in global models of disks with vertical flux in the non-relativistic regime \citep{Zhu2018}.  The radial shear of the vertical magnetic field between midplane and corona is evident in the blue magnetic field streamlines plotted for models E08-a9, E08-a3, and E07-a3-DL in \autoref{fig:profile2d}.

Due to the strongly magnetized corona, there is no evidence that these gas pressure supported thin disks undergo thermal instability.  However, some sort of thermal runaway may have contributed to the formation of this state in the first place.  At these low accretion rates cooling is primarily due to radiative diffusion (as shown in the lower right panel in \autoref{fig:hori1d_compare}). 

As the mass accretion rate is decreased to sub-Eddington levels, the geometric size and optical depth of the magnetic corona decreases which makes the photosphere closer to the surface of the thermal disk component.  This may have important implications for understanding state changes in BH X-ray binaries, and in particular the emergence of the soft state.

\subsubsection{Magnetically elevated disk}
\label{sec:magp_disk}

In some cases, models that begin with a double-loop magnetic field (no vertical flux at the midplane) settle into a disk which is magnetically-dominated everywhere \citep{Begelman2024}, with no density enhancement at the midplane (see the right panels in \autoref{fig:disk_Bfield}).  The disk is supported primarily by both magnetic and radiation pressure throughout, with the former dominating.  Angular momentum is transported outward through Maxwell stress and turbulent Reynolds stress.  No evidence of thermal instability is evident, although given the strong magnetic pressure support, none is expected \citep{Sadowski2016}.

\subsection{Comparison to Previous Models}
\label{sec:previous-work}

While there are numerous previous models of radiation-dominated accretion reported in the literature, direct comparison to our results is complicated by a number of factors, including (1) different models start with different initial conditions, either geometrically thicker disks or disks with large amplitude net vertical magnetic flux than adopted here (2) other models are often evolved for many dynamical times before radiation is included in the dynamics, (3) the radiation transport in other models is treated with various approximations, from FLD to the M1 method.  Nevertheless, many of the general trends reported here are similar to previous models, including the emergence of thick disks with powerful outflows in the super-Eddington regime, and the emergence of magnetically dominated coronae in the near- and sub-Eddington regime.

One way that our models differ from previous work is in the lack of emergence of a MAD state, despite long-term evolution of the flow.  Previous authors who have studied the MAD state in radiation-dominated disk models began with additional vertical magnetic field in the initial conditions compared to our single-loop models.  This may be essential for achieving the MAD state in thin disks.

A second and perhaps more important difference between our results and previous models is the measured radiative efficiency of the disk.  We find efficiencies of at most a few percent at the lowest Eddington ratios, and this drops to $\sim 0.5$\% for super-Eddington flows.  Other authors have reported much larger values, up to almost $100$\% \citep{McKinney2017, Fragile2023, Curd2023}.  There are several possible reasons for this discrepancy.  Firstly, the highest measured efficiencies are for disks that reach the MAD state, whereas our models do not.  Secondly, the efficiencies are measured at different radii by different authors, and may include radiation emitted by the disk if it is initially hot (geometrically thick) rather than produced by accretion.  Finally, it is possible that radiative efficiencies may depend on the radiation transport model adopted.  Therefore, examining radiation-dominated models of MAD accretion flows using the full transport algorithms developed here is an important direction for future work. 

Recently, \citet{Fragile2025} have reported that radiation GRMHD models of super-Eddington accretion, initialized with a generalized analytical thin disk model, obey the Eddington limit, that is  $\dot{M} \lesssim\dot{M}_{\mathrm{Edd}}$ in all of their calculations regardless of the initial conditions, a result which is quite different from those reported here.  These authors emphasize the importance of studying flows in which the trapping radius is small compared to the overall size of the disk \citep[e.g.][]{Kitaki2021}.  However, based on the measured location of the trapping radius (see \autoref{fig:hori1d_compare}), this condition is satisfied in our models.  Thus the origin of the quite different outcomes between these calculations is uncertain and warrants further investigation.

Finally, we can also compare our results computed using the GRMHD extension of our radiation transport algorithms to our previous non-relativistic models reported in a series of papers \citep{Jiang2014b, Jiang2019ApJ880, Jiang2019ApJ885, HuangJiang2023}.  Overall, our results are remarkably consistent, for example, we find the radiative efficiencies of the disks are within a factor of a few, we find turbulent convection dominates radiative diffusion for cooling of the disk, and we find the vertical scale height of the magnetic field is much larger than the density scale height resulting in magnetically dominated coronae at low mass accretion rates.  However, there also are ways in which are models differ.  For example, the GRMHD models reported here capture the formation of a relativistic jet for spinning black holes, which contributes to the overall energy output rate and cooling from the inner disk.  Moreover, our models also capture the dynamics of the gas in the plunging region, allowing comparison to analytic models \citep{Mummery2024}. We also find differences in the detailed structure of the inner disk between our two sets of calculations, for example we do not see the emergence of strong global spiral waves in the inner disk reported in the AGN models of \citet{Jiang2019ApJ880}.  However this is consistent with earlier super-Eddington models of stellar mass black holes \citep{Jiang2014b}. 
Our earlier models also lacked the resolution to capture the formation of a thin ($H/r \approx 0.02$), gas-pressure dominated disk (see \autoref{fig:disk_Bfield}).

\section{Observational Implications} 
\label{sec:obs_implication}

In this section, we highlight some of the implications of our results for the interpretation of observations of a variety of systems, including X-ray binaries and ultraluminous X-ray sources (ULXs). 
Moreover, although our models have been computed using opacities appropriate to accretion flows around stellar-mass black holes, we discuss possible applications to little red dots (LRDs), as similar radiation-driven dynamics are expected in the inner disk region where the hot gas is fully ionized and Thomson scattering dominates.

\subsection{Radiative efficiency and variability}
\label{sec:EffVar}

\begin{figure}
    \centering
    \includegraphics[width=\columnwidth]{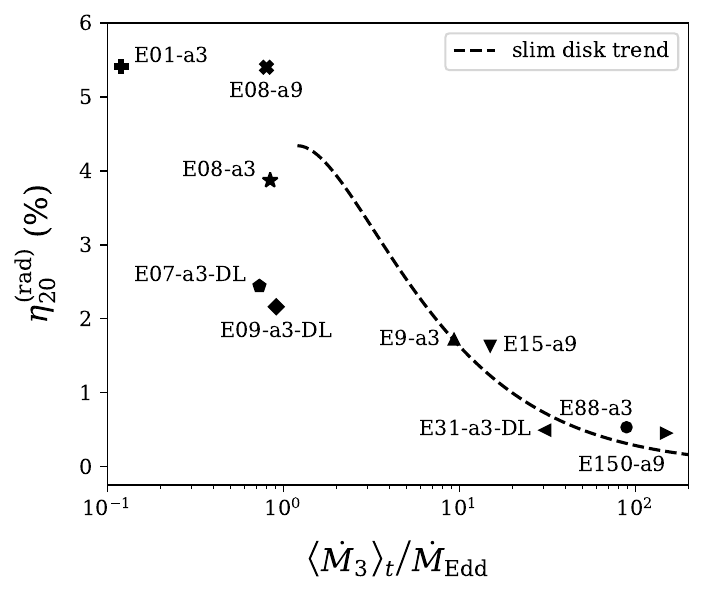}
    \caption{
    Radiation efficiency versus mass accretion rate for all models. The dashed line shows the trend predicted by analytical models of super-Eddington accretion (e.g. equation~6 in \citealt{Poutanen2007}), normalized to match our model E9-a3. 
    }
    \label{fig:rad_eff}
\end{figure}

In \autoref{fig:rad_eff}, we show the radiation efficiency as a function of the mass accretion rate for all our models.  The radiation efficiency includes emission from within $r=20r_g$, while the time-averaged mass accretion rate is evaluated at $r=3r_g$.  A strong decrease in efficiency in the super-Eddington regime is clearly evident.  This trend is remarkably consistent with expectations from the slim disc model \citep{Abramowicz88} shown by the dashed line.  In this model, the trapping radius increases with accretion rate \citep{Begelman1979}, leading to the observed decline. Recently, \citet{Poutanen2007} has shown that incorporating the effects of outflows in the super-Eddington regime (see their equation~19) can also lower the efficiency, in agreement with observational data. The low radiative efficiency of super-Eddington models suggests that observationally these sources would never appear with highly super-Eddington bolometric luminosities.  
The highest radiation efficiency among our models occurs in the sub-Eddington regime, although even then it never exceeds $\sim 5$\%.  Note that models initialized with the double-loop magnetic configuration generally exhibit lower radiation efficiency by a factor of two.  As discussed earlier, these results differ from some previous results, where radiative efficiencies of nearly $100$\% are reported for both near- and super-Eddington regimes \citep{McKinney2017, Curd2023}.

\begin{figure}
    \centering
    \includegraphics[width=\columnwidth]{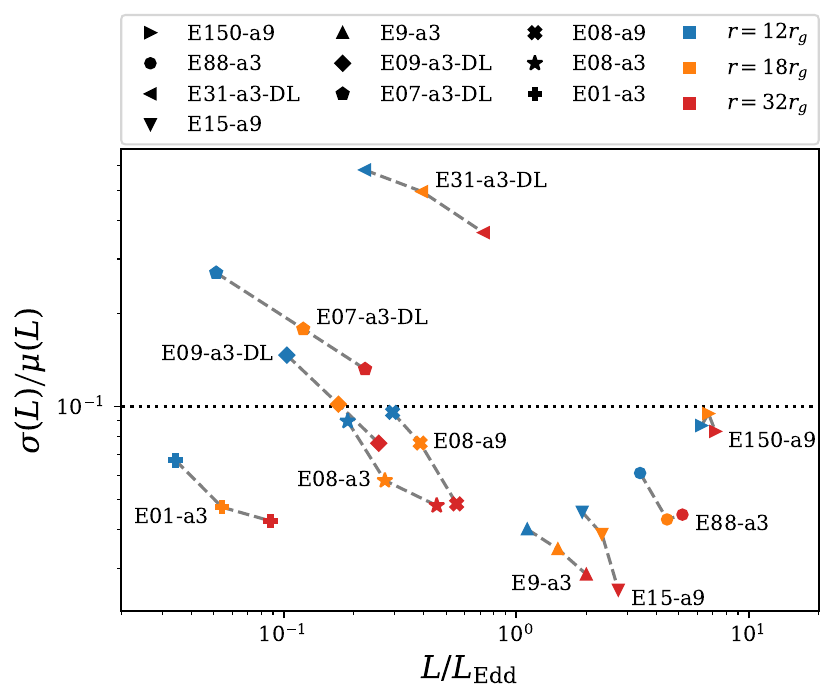}
    \caption{Amplitude of luminosity variability (defined as the RMS deviation divided by the mean) versus total output radiation luminosity for all our models. Both quantities are measured at three different radial locations denoted by the blue, orange, and red symbols.  The dotted horizontal line marks a 10\% variability threshold.  Due to the very low radiative efficiencies, even the most super-Eddington accretion flows only emit $\lesssim 10 L_{\mathrm{Edd}}$. Models with the double-loop magnetic field geometry produce the highest variability. 
    }
    \label{fig:var_compare}
\end{figure}

In \autoref{fig:var_compare}, we plot the amplitude of luminosity variability (defined as the RMS deviation divided by the mean) versus total output radiation luminosity for all our models.  Both quantities are evaluated within a time window of approximately $10^4 r_g/c$ (0.5~s for a $10 M_\odot$ black hole) for each simulation.  Due to I/O limitations, we have output 3D data at relatively coarse time intervals ranging between $100r_g/c$ (5~ms) and $250r_g/c$ (125~ms) depending on the model.  We remove low-frequency variation by subtracting a second-order polynomial fit, as low-frequency modes are not sufficiently sampled by our data.  We measure variability at three different radii ($r=12$, $18$, and $32r_g$), all of which are within the region where most of the X-ray luminosity is generated.  

\autoref{fig:var_compare} shows that models that begin with a single field loop geometry produce significantly less variability ($\lesssim 10$\%) than those that begin with a double-loop geometry, with the smallest variability ($\sim 3$\%) associated with the model E9-a3 which is moderately super-Eddington.  At low accretion rates, the photosphere is close to the disk midplane, and variability is introduced by turbulence in the magnetically dominated corona.  At high accretion rates, a radiation-supported geometrically thick disk emerges, and the photosphere is located closer to the funnel region where variability may be introduced through outflows.  The near-Eddington cases seem to lie between these regimes.

The three models that begin with a double-loop magnetic field geometry produce the highest variability across a factor of $30\times$ in mass accretion rate.  A common feature of these models in the near-Eddington regime is a highly turbulent magnetic envelope (see \autoref{fig:disk_Bfield}) in which the disk photosphere is embedded, which is likely the origin of the stronger variability.  In the super-Eddington regime, the double-loop configuration fails to produce strong jets to evacuate the funnel region, which also leads to a more turbulent photosphere.

Finally, \autoref{fig:var_compare} also demonstrates that in each model variability generally decreases as the radii is increased from $12r_g$ (blue symbols) to $32r_g$ (red symbols).  This effect is largest at the lowest accretion rates.  This suggests that observed variability may be wavelength dependent.

\subsection{Beaming factors}
\label{sec:beaming}

\begin{figure}
    \centering
    \includegraphics[width=\columnwidth]{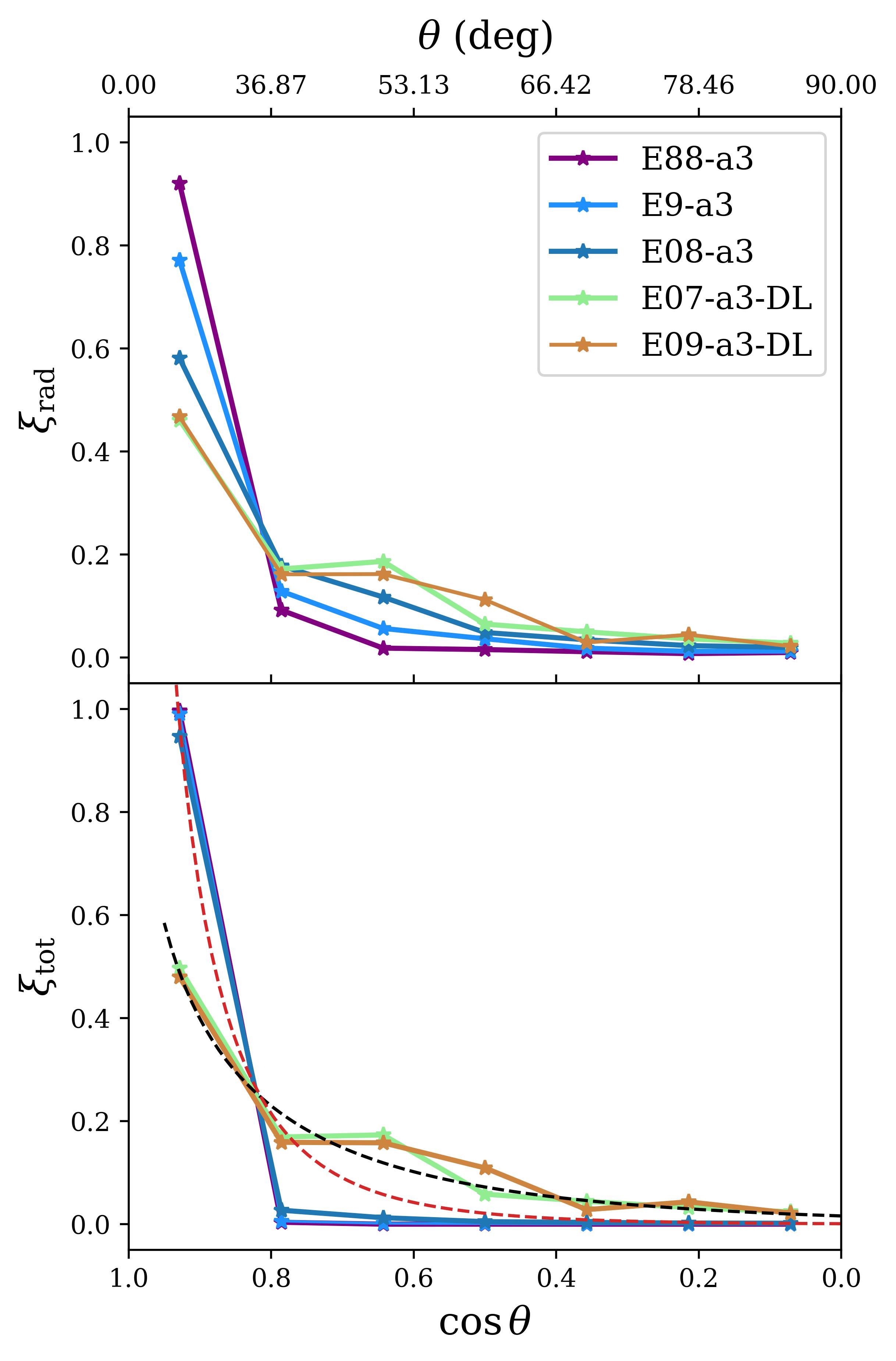}
    \caption{Angular distribution of outgoing radiation field (upper panel) and total energy outflow (lower panel) at $r=1024r_g$ measured by the beaming factors (see equation~\ref{eq:beaming_factor} for definitions) for selected models. In most cases, both the radiation field and total energy outflow are strongly beamed. In the lower panel, the dashed lines are fitted using equation~7 from \citet{Sadowski2015c} normalized to match models E88-a3 (red line), and E07-a3-DL (black line) respectively.   
    }
    \label{fig:beam}
\end{figure}

In \autoref{fig:beam}, we present the angular distribution of the beaming factors for both radiation and total (radiation plus kinetic) luminosities, measured at $r=1024r_g$.  To minimize ray effects associated with the finite number of solid angles used to discretize the specific intensity, the outgoing energy flux is averaged over $N_l = 7$ bins over the range $0^{\circ} \leq \theta \leq 90^{\circ}$. The resulting definitions of the beaming factors are given by: 
\begin{subequations}    
\begin{align}
    & \xi_{\mathrm{rad}} = 
    \frac{\displaystyle\int_{\theta_l}^{\theta_{l+1}}\left<-R^r_{\ t}\right>_{\phi,t}^{(\mathrm{hemi})}\sqrt{-g}d\theta}{\displaystyle\int_{0}^{\pi/2}\left<-R^r_{\ t}\right>_{\phi,t}^{(\mathrm{hemi})}\sqrt{-g}d\theta}
    \ ,
    \\
    & \xi_{\mathrm{tot}} = 
    \frac{\displaystyle\int_{\theta_l}^{\theta_{l+1}}\left<-T^r_{\ t}-R^r_{\ t}-\rho u^r\right>_{\phi,t}^{(\mathrm{hemi})}\sqrt{-g}d\theta}{\displaystyle\int_{0}^{\pi/2}\left<-T^r_{\ t}-R^r_{\ t}-\rho u^r\right>_{\phi,t}^{(\mathrm{hemi})}\sqrt{-g}d\theta}
    \ ,
\end{align}
\label{eq:beaming_factor}
\end{subequations}
where $l=(1,N_l)$ and the $\theta_l$ are selected such that each bin has equal solid angle.  Note that we average the azimuthally and time-averaged radiation and total energy fluxes over both the northern and southern hemispheres.  Within each bin, the energy flux is set to zero if the corresponding region lies within the disk body.  Using these definitions, the beaming factors quantify the fractional luminosity distributed across each solid angle bin, such that their sum over all bins equals unity.  The black and red dashed lines in the lower panel show the results of applying the fitting function from \citet{Sadowski2015c}, using characteristic jet beaming angles of $20^{\circ}$ and $10^{\circ}$, respectively, and normalized to match our models E07-a3-DL and E88-a3.

It is clear from the upper panel of \autoref{fig:beam} that the radiation field is highly beamed for super-Eddington accretion models, largely because the photosphere is restricted to a narrow funnel region around the $z$-axis (see the left column in \autoref{fig:profile2d}).  In near-Eddington models, the disk is much thinner and the photosphere is located closer to the equatorial plane, resulting in less beaming.  The lowest beaming factors are associated with models that initialized with double-loop magnetic fields.  In these cases, there is a very weak jet, leading to even less collimation in the inner regions.

The collimation of the radiation field can also be observed in the streamlines of the radiation flux plotted in the right panels of \autoref{fig:profile2d} for selected models.  In particular, very strong collimation is reflected by the narrow cone of nearly vertical streamlines in the super-Eddington models in the left column.  Much broader streamlines are evident in the optically thin wind region above the photosphere in the near-Eddington models plotted in the middle column. 

In the lower panel of \autoref{fig:beam}, the beaming of the total energy outflow shows a strong dependence on magnetic field topology: single-loop configurations with strong jets formed are highly beamed near the pole, while double-loop configurations with weak or absent jets exhibit significantly less beaming. 

\subsection{Spectral differences}
\label{sec:SpecDiff}

Our models solve the frequency-integrated radiation transfer equation using gray opacities and therefore the outgoing radiation field does not contain information regarding the spectra.  By post-processing snapshots of our simulations using frequency-dependent transport methods and realistic opacities, we plan to present spectra of our models in the future.  Nevertheless, it is still possible to predict some aspects of the spectra resulting from our models.  In particular, by calculating the ratio between the gas temperature to the effective temperature of the radiation field along the surface of the scattering photosphere, it is possible to estimate how strongly the emerging spectra may deviate from a blackbody.  The effective temperature $T_{\mathrm{eff}}$ is calculated based on the outgoing fluid-frame radiation flux as   
\begin{equation}
    T_{\mathrm{eff}} = \left(\frac{\bar{F}_r^{\perp}}{\sigma_{\mathrm{SB}}}\right)^{0.25}
    \ ,
\end{equation}
where the outgoing flux $\bar{F}_r^{\perp}$ is perpendicular to the surface of the scattering photosphere and $\sigma_{\mathrm{SB}}$ is the Stefan-Boltzmann constant. For near-blackbody emission, one expects $T_{\mathrm{gas}} \simeq T_{\mathrm{eff}}$, while $T_{\mathrm{gas}} \gg T_{\mathrm{eff}}$ implies a potentially much harder spectrum.  

\begin{figure}
    \centering
    \includegraphics[width=\columnwidth]{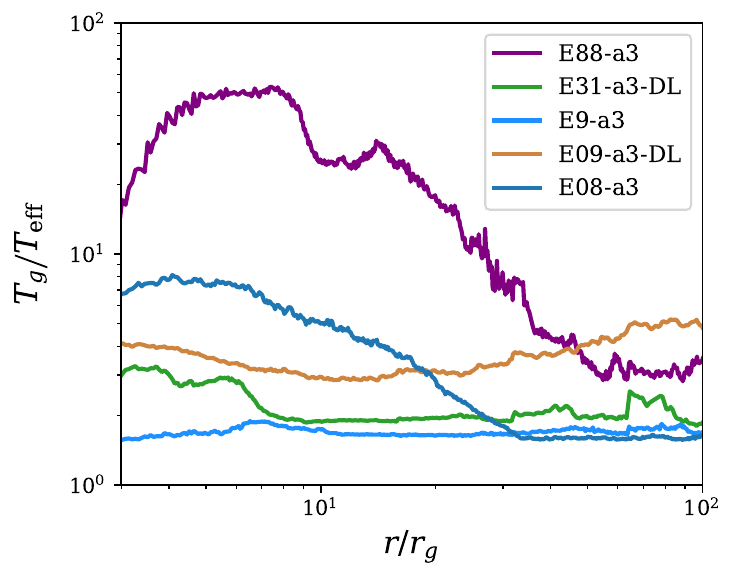}
    \caption{Ratio of gas temperature to effective temperature of the radiation field along the scattering photosphere for selected models. 
    }
    \label{fig:temp_ratio}
\end{figure}

We compute the temperature ratio using time- and azimuthally averaged quantities.  \autoref{fig:temp_ratio} plots this ratio for five models with low spin and mass accretion rates ranging from near- to highly super-Eddington.  These models produce a wide range of behavior.  The moderately super-Eddington model E9-a3 is likely to produce a spectrum closest to a blackbody due to the near equilibrium between the gas and effective temperatures at the photosphere in this model.  In this model, the photosphere is well above the regions where dissipation of turbulence heats the gas.  On the other hand, in the highly super-Eddington model E88-a3, spectral hardening can become significant.  In this case the photosphere is close to the low density and highly magnetized funnel region, so that magnetic dissipation in regions of low density contribute to heating the gas. Consequently, the system enters into a photon-starved regime, where the free-free emission is capable of providing the necessary cooling only if $T_{\mathrm{gas}} > T_{\mathrm{eff}}$, which should lead to significant spectral hardening, even with Compton scattering contributing to the cooling \citep{Davis2019}.

Note that run E31-a3-DL, a super-Eddington model that begins with a double-loop magnetic configuration has a remarkable thermal spectrum.  Runs that lack net magnetic flux at the midplane produce weak jets.  In this model, the jet is too weak to clear the funnel region; instead, the funnel is filled by optically thick, radiation-driven outflows from the disk, resulting in a relatively thermal spectrum emerging from the photosphere.

For the two near-Eddington models (E08-a3 and E09-a3-DL), magnetic dissipation above the photosphere can produce a harder spectrum compared to E9-a3. Since much of the dissipation occurs in regions that are optically thick to electron scattering, most of the emission is likely in the saturated Compton limit, giving rise to Wien-like spectra that can be approximated as color-corrected (or diluted) blackbody spectra with a color correction $f \sim T_{\mathrm{gas}}/ T_{\mathrm{eff}}$ \citep{ShimuraTakahara1995}.

\subsection{Black hole X-ray binaries in the soft state}
\label{sec:xraybinaries}

Black hole X-ray binaries in the soft state are typically associated with near- or sub-Eddington accretion in a radio-quiescent state and exhibit a low luminosity variability (e.g., see \citealt{Kalemci2022} for details). These observational properties pose a fundamental challenge for theoretical models.  In particular, models with weak poloidal magnetic field (e.g. those that begin with a double-loop initial field) are favored since they lack powerful relativistic jets that produce radio emission.  However, as seen from \autoref{fig:var_compare}, these models produce greater variability in luminosity. On the other hand, while a net poloidal field may create a relatively stable thermal disk with smaller luminosity variations, it also promotes strong jet formation, making the system radio-loud.

We find that models with the lowest variability in agreement with observations are all super-Eddington (e.g. E9-a3 and E15-a9), and likely not applicable to these sources.  In near-Eddington simulations, a geometrically thin thermal disk surrounded by an optically thick magnetic envelopes develops, producing variability amplitudes that are too large (e.g., E08-a3 and E08-a9).  Thus, our best-fit model lies in the sub-Eddington regime, in which the magnetic envelope surrounding the thermal-pressure supported disk is found to be more optically thin (e.g., E01-a3) and therefore produces less variability. Moreover, radial transport of poloidal magnetic flux by the thin disk is small in this model, leading to a low amplitude of magnetic flux at the horizon and a weak (or absent) jet.  In fact, we find that radiation pressure can elevate inflows above the thin disk, triggering magnetic reconnection at the poles and completely suppressing jet formation.  Further application of our sub-Eddington models to X-ray binaries will be discussed in Paper IV.  

\subsection{Ultraluminous X-ray sources (ULXs)}
\label{sec:ulx}

Ultraluminous X-ray sources (see \citealt{Kaaret2017} and \citealt{King2023} for reviews) are point-like, non-nuclear objects with X-ray luminosities exceeding $10^{39}~\mathrm{erg~s^{-1}}$, corresponding to the Eddington luminosity of a 10-solar-mass black hole accretor, assuming isotropic radiation emission.  Most ULXs are interpreted as systems undergoing super-Eddington accretion onto a stellar-mass object, either a neutron star exhibiting coherent pulsations (e.g., \citealt{Bachetti2014}) or a stellar-mass black hole characterized by spectral features \citep{Gladstone2009}.  

Recently, the Imaging X-ray Polarimetry Explorer (IXPE; \citealt{Weisskopf2022}) identified Cyg X-3 as a galactic ULX, where the central black hole undergoes super-Eddington accretion and exhibits a high degree of X-ray polarization \citep{Veledina2024, Veledina2024b}.  A funnel region is necessary to explain the observed high polarization degree, as it likely arises from the reflection off the inner wall of the funnel, aligned with the direction of the observed radio ejections.  During the spectral transition from the hard to soft state, Cyg X-3 shows a decrease in both accretion rate and polarization degree, while its radio flux may either increase or decrease (see \citealt{Veledina2024} for details). 

In our near-Eddington runs, the emergent radiation from the inner disk can be moderately beamed, as shown by models E08-a3 and E08-a9 in \autoref{fig:beam}, where the apparent luminosity can reach the ULX regime at appropriate viewing angles.  In our super-Eddington runs, a conical funnel naturally forms when the thermal expansion of the accretion disk is halted by the jet near the black hole, as illustrated in the first column of \autoref{fig:disk_Mdot}.  The narrow opening angle of this funnel allows only a small fraction of radiation from the inner disk to escape directly, which results in beamed outgoing radiation and low radiation efficiency, consistent with the scenario proposed by, e.g., \citet{King2024}.  As the accretion rate increases in the super-Eddington regime, the funnel opening angle narrows slightly, further enhancing the beaming effect.  Once the accretion rate becomes sufficiently high, the system can even enter a photon-starved regime \citep{Davis2019}, where spectral hardening becomes significant, as demonstrated by the E88-a3 case in \autoref{fig:temp_ratio}.

It is interesting to compare the observed properties of Cyg X-3 with our super-Eddington runs, where the properties of the flow closely align with the observed state transitions reported by \citet{Veledina2024, Veledina2024b}.  Similar spectral transition are also seen in NGC~253~X-2, NGC~1313~ULX-2, and NGC~4190~ULX1 \citep{Kajava2009,Ghosh2021}.  It is worth noting that some black hole ULXs exhibit the opposite trend, where the spectrum softens as the luminosity increases \citep{Kajava2009}.  The changes in disk configuration from the near-Eddington to super-Eddington regime observed in our simulations provide a plausible explanation for this behavior.  In the near-Eddington case, the presence of a magnetic corona can produce a harder spectrum (see E08-a3 in \autoref{fig:temp_ratio}), whereas in the super-Eddington case, before entering the photon-starved regime, the radiation-supported disk tends to produce a softer spectrum (see E9-a3 in \autoref{fig:temp_ratio}). 

The consistency between these observed phenomenology and our simulation results reinforces the critical role of accretion structure and outflow geometry, shaped by radiation dynamics, in interpreting X-ray polarization and spectral transitions across black hole ULX populations.

\subsection{SS 433}
\label{sec:ss433}

SS 443 \citep{Fabrika2004} is another well studied galactic X-ray binary, which is noted for its precessing jets. These jets have a velocity of $\sim 0.26 c$, emit strongly in H$\alpha$, are collimated within a cone with an opening angle of about $40^\circ$, and precess on a 162 day orbit. It also shows broad absorption lines indicative of a strong equatorial outflows with velocities as high as $\sim 1300 \; \rm km/s$. Even though its X-ray luminosity is only $\sim 10^{36} \; \rm erg/s$, it is often modeled as a ULX viewed edge on, with inner X-ray emitting regions obscured because we view the source at high inclination \citep{Begelman2006, Middleton2021}.

Our simulations do not show any evidence of jet precession and the jet velocities seen in Figure~\ref{fig:ut_jet} are much higher than the $0.26 c$ found in SS 433. This, however, only reflects the results from simulations where the velocity streamlines trace back to the strongly magnetized inner regions, which excludes the double-loop simulations (see \autoref{tab:sim_result}). The outflow velocities along the polar axis in these simulations are similar to the SS 443, with $u^t \sim 0.1$ for a cone with opening angle $\sim 30^\circ$. This imply that the modest jet velocity in SS 433 is primarily the result of radiation pressure, with poloidal magnetic fields being too weak to launch a powerful MHD jet. Our simulations also launches an outflow of lower velocity gas that is qualitatively consistent with the equatorial flow observed in SS 443. The velocities in our simulations tend to be somewhat higher (e.g., see Figure~\ref{fig:glb_structure}), but these are likely impacted by the ongoing relaxation of the torus at large radius.  In Paper II, we will present a detailed analysis of the super-Eddington model E31-a3-DL, which develops a weak jet partially powered by radiation and likely represents a physical scenario analogous to SS~433.  This also provides an opportunity to investigate how system properties depend on magnetic topology in ways that can be tested through observations.

\subsection{Little red dots (LRDs)}
\label{sec:lrds}

\begin{figure}
    \centering
    \includegraphics[width=\columnwidth]{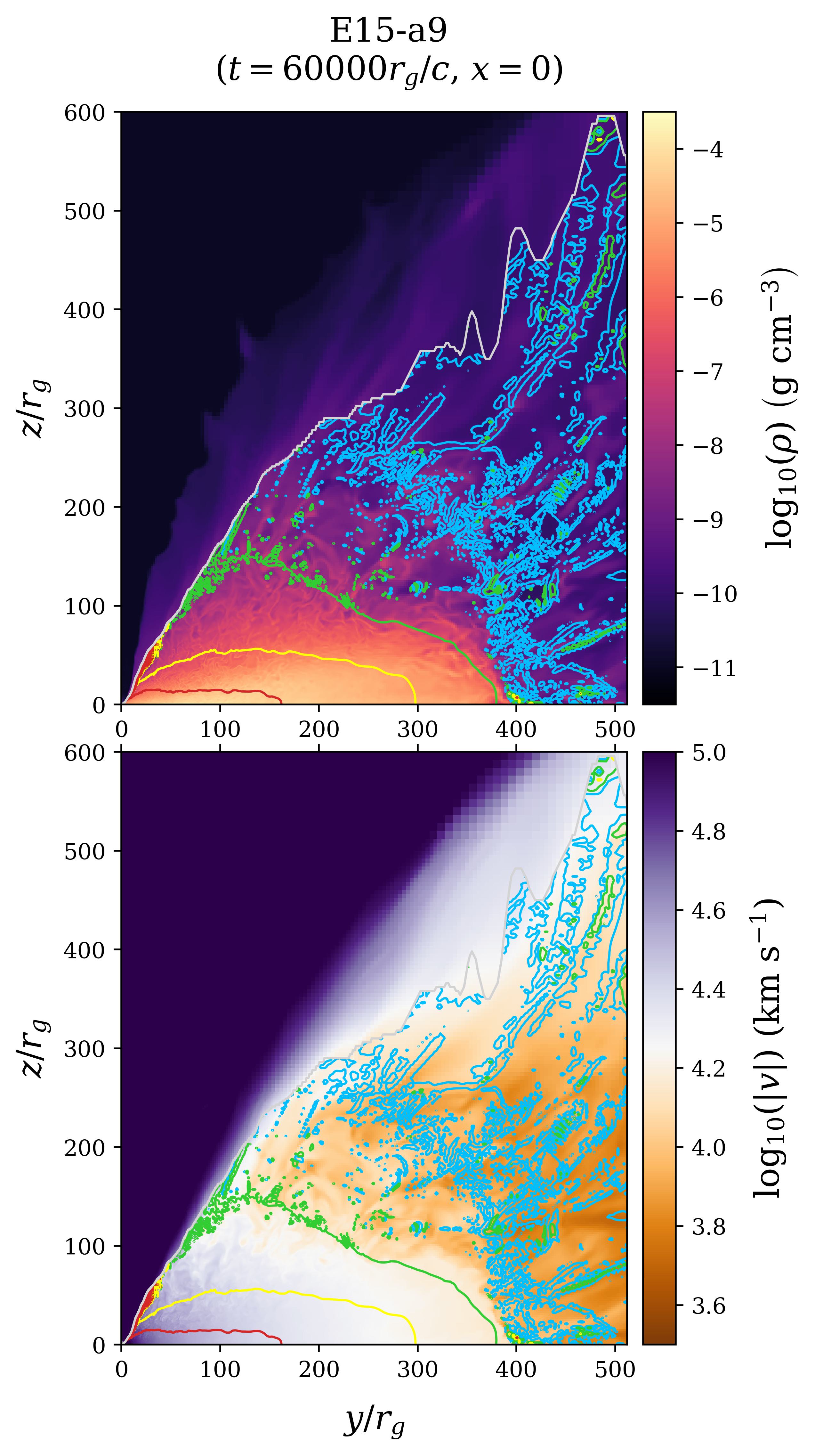}
    \caption{Density (top panel) and magnitude of outflow velocity (bottom panel) of a super-Eddington accretion disk on large scales, relevant for interpreting observations of LRDs.  The grey lines represent the scattering photosphere.  The red, yellow, green, and blue lines represent the isothermal contours of the logarithmic gas temperature in Kelvin, with values of 7.3, 7.0, 6.6, and 6.3, respectively.
    }
    \label{fig:glb_structure}
\end{figure}

The James Webb Space Telescope (JWST) has discovered a population of high-redshift compact red sources, characterized by broad emission lines (FWHM $> 2000~\mathrm{km~s^{-1}}$), suggesting that active galactic nuclei (AGN) are strong candidates for these sources \citep{Matthee2024, Greene2024, Wang2024}.  Despite the different scales of central black hole mass explored in this paper, the radiation-driven dynamics remain similar in the super-Eddington regime, as the inner disk is hot and fully ionized, with Thomson scattering dominating the opacity.  Therefore, our super-Eddington accretion simulations still provide qualitative insights into these objects.

To illustrate the large-scale structure of our super-Eddington accretion models, \autoref{fig:glb_structure} plots color images of the density and magnitude of the outflow velocity overlaid by contours of the temperature for model E15-a9 with $\dot{M} = 15 \dot{M}_{\mathrm{Edd}}$.  The image clearly shows the geometrically-thick, radiation-supported disk in the inner regions, with a slow and dense equatorial wind that subtends a large angle.
As discussed below, the structure revealed in the images suggests explanations of two of the most puzzling characteristics of the little red dots (LRDs).

The first puzzle is that all the spectroscopically confirmed LRDs are underluminous in X-ray (e.g., \citealt{Kocevski2023, Maiolino2023, Wang2024, Greene2024, Matthee2024, Ananna2024, Yue2024, Sacchi2025}).  The geometrically thick accretion disk formed in super-Eddington accretion flows evident in \autoref{fig:glb_structure} suggest an obvious explanation: when a strong jet forms and evacuates the polar funnel, radiation generated in the inner disk can escape only through this narrow, optically thin region.  That is, radiation from the inner disk in super-Eddington models is highly beamed (see \autoref{sec:beaming}), an idea that has also been suggested by other authors (e.g., \citealt{King2024b}).  In a super-Eddington model with only a weak jet (E31-a3-DL), the funnel remains uncleared; instead, radiation-driven outflows fill this region, further preventing X-ray photons from escaping the system.  This further provides an additional explanation for the underdetection of X-ray emission in such systems, as high-energy photons can escape only if a strong jet formed, which is not guaranteed in all cases (e.g. low black hole spin or insufficient net vertical magnetic flux). 

Second, many LRD spectra show a change in continuum slope near $\lambda \approx 3600$~\AA, sometimes with a distinct break \citep{Setton2024}.  \citet{Wang2024,Wang2024b} interpret this as a Balmer break originated from old stellar populations, implying unexpectedly early star formation and extreme stellar masses at $z > 6$. This tension motivates alternative explanations. \citet{Inayoshi2024} propose a pure AGN origin where gas between the broad and narrow line regions absorbs AGN emission, producing a sharp cutoff near the Balmer limit.  Our simulations provide a more physical scenario: radiation-driven outflows travels a large distance, gradually cools radiatively, and decelerate due to gravity, with some material falling back to form a cold layer above the hot disk.  This creates strong temperature contrasts (see \autoref{fig:glb_structure}), enabling the formation of a smooth break around the Balmer limit when scaled to $10^4$ -- $10^5$~K for typical AGN conditions.  Outflow velocities decelerate to a few thousand $\mathrm{km~s^{-1}}$, as shown in the lower panel of \autoref{fig:glb_structure}, consistent with observations \citep{Wang2024b}.  This scenario further suggests that sources exhibiting such absorption features are less likely to show X-ray emission, as the cold, optically thick gas layer obscures most of the regions emitting hot, high-energy photons.

\section{Conclusions} 
\label{sec:conclusion}

In this paper we have presented an overview of the results from a survey of radiation-dominated accretion flows around black holes with a wide range of different mass accretion rates and spins, calculated using new full-transport algorithms for radiation GRMHD.  These models can be considered an extension of our previous non-relativistic calculations \citep[e.g.][]{Jiang2014b, HuangJiang2023} to full GR.

While our models use opacities appropriate for stellar-mass black holes, it is likely that many general features of our results will also apply to accretion onto supermassive black holes as well.  Our main results are the following:
\begin{itemize}
    \item Turbulence produced by the MRI drives mass accretion at rate that vary from highly sub- to super-Eddington depending on the initial mass surface density.  There is no Eddington limit to accretion. 
    \item In the super-Eddington regime, accretion produces a geometrically thick, radiation pressure supported disk with powerful outflows, with mass outflow rates that are comparable to the mass accretion rate.
    \item At near- and sub-Eddington rates, the structure of the disk depends on whether it contains a net vertical flux at the midplane.  Models with net flux produce a thin ($H/r \sim 0.02$), dense gas layer at the midplane with a magnetically dominated corona.  Models without remain magnetically dominated everywhere.
    \item The overall radiative efficiency of accretion drops dramatically as the mass accretion rate is increased, from a few percent at sub-Eddington rates to $\lesssim 0.5$\% for highly super-Eddington.
    \item Radiation-driven winds are formed in all cases.  Such winds carry as much energy as the emergent radiation field in the super-Eddington regime, but significantly less in the sub-Eddington regime.
    \item Vertical advection of radiation energy by turbulence dominates radiation diffusion and therefore cooling of the disk in the radiation-dominated, super-Eddington regime, a result which is consistent with non-relativistic models of \citet{Jiang2014b}.
    \item Relativistic jets are produced in models that begin with net vertical magnetic flux at the midplane and a rapidly spinning black hole.
    \item None of the models we study enter the MAD regime.  Previous models that achieve the MAD state in thin disks require the addition of significant net vertical flux to the inner regions of the disk.
\end{itemize}

Our results have important implications for the interpretation of observations of a variety of astrophysical systems.  For example:
\begin{itemize}
    \item The very low radiative efficiency we measure for super-Eddington accretion implies such sources should not produce intrinsically highly super-Eddington luminosities. 
    \item The emergent radiation is highly beamed in super-Eddington disks, which enables observations of highly super-Eddington fluxes at favorable viewing angles. 
    \item The variability of the emergent luminosity on timescales below $10^{4}r_g/c$ spans a wide range depending on the initial magnetic field geometry (and therefore resulting structure of the disk).  Typically, it is $\lesssim 10$\% in the most super-Eddington cases. 
\end{itemize}
In \autoref{sec:obs_implication}, we discuss the application of our results to specific astrophysical systems, including the soft state of X-ray binaries, ULXs and SS433, and LRDs.

In this summary paper, we have only provided an overview of our results. Papers II through IV in this series will present a more in-depth analysis of super-Eddington, near-Eddington, and sub-Eddington flow regimes, respectively. 

A potential limitation of our models is that they begin with a tightly bound torus of plasma whose density and pressure maximum is at $r=58r_g$.  From these initial conditions we find that highly super-Eddington accretion can be established and maintained.  However, there are important questions about how plasma is fed from beyond the Bondi radius to the inner regions, and whether radiation and mechanical feedback from outflows can limit this feeding.  We find highly super-Eddington flows are strongly beamed (see \autoref{fig:beam}) consistent with previous results \citep[e.g.,][]{Sadowski2015c}, thus it is highly likely that highly super-Eddington accretion can be maintained through equatorial inflows \citep[see][]{Kaaz2025}.  Nevertheless, further studies of mass inflow and feedback onto accreting black holes in realistic conditions is important \citep[e.g.,][]{Guo2024}. 

Finally, there are a number of promising directions for future investigation.  One possibility is to use Monte-Carlo radiation transport methods to compute detailed spectra from snapshots of our models.  Another direction involves incorporating a fully relativistic frequency dependent transport algorithm based on \citet{Jiang2022}, which could support frequency-dependent radiation GRMHD models of AGN disks in the sub-Eddington regime, where accurate opacity treatment is particularly important.  A version of the M1 approximate radiation transport scheme has also been implemented in the \texttt{AthenaK} code, providing opportunities for direct comparison with solutions computed using full-transport methods.  In addition, adapting the cyclic-zoom algorithm of \citet{Guo2025} to radiation GRMHD could enable the evolution of radiation-dominated AGN disks to be modeled over very large length- and time-scales.  Finally, the adoption of novel tensor network methods to accelerate radiation transport calculations \citep{TensorTrain2025} could make it feasible to employ larger numbers of angular and frequency bins in future models, improving the accuracy of radiation-dominated disk simulations.

\begin{acknowledgments}

We thank the anonymous referee for helpful remarks that improve this paper.  We also thank Omer Blaes, Bingjie Wang, Charles Gammie, George Wong, Chris Done, Julian Krolik, and Matthew Middleton for useful discussions. This work was supported by the Schmidt Futures Fund, NASA TCAN grant 80NSSC21K0496, and the Simons Foundation.  The analysis made significant use of the following packages: NumPy \citep{Harris2020}, SciPy \citep{Virtanen2020}, and Matplotlib \citep{Hunter2007}.

An award for computer time was provided by the U.S. Department of Energy's (DOE) Innovative and Novel Computational Impact on Theory and Experiment (INCITE) Program. This research used supporting resources at the Argonne and the Oak Ridge Leadership Computing Facilities. The Argonne Leadership Computing Facility at Argonne National Laboratory is supported by the Office of Science of the U.S. DOE under Contract No. DE-AC02-06CH11357. The Oak Ridge Leadership Computing Facility at the Oak Ridge National Laboratory is supported by the Office of Science of the U.S. DOE under Contract No. DE-AC05-00OR22725. We thank Vassilios Mewes and Kyle Felker for support on these facilities.

Research presented in this article was supported by the Laboratory Directed Research and Development program of Los Alamos National Laboratory under project number 20220087DR.

This work has been assigned a document release number LA-UR-25-25209.

\end{acknowledgments}

\newpage
\appendix
\section{Definitions of timescales used in diagnostics}
\label{appendix:timescales}

The timescales plotted in \autoref{fig:hori1d_compare} are defined as follows.  The inflow time is given by
\begin{subequations}
\begin{align}
    t_{\mathrm{inflow}} &= -r/\langle v^r \rangle
    \ ,
    \\
    \langle v^r \rangle &= \frac{\displaystyle\int_{\mathrm{disk}} \left<\left(\rho + \frac{\gamma}{\gamma-1}P_g + b^{\lambda}b_{\lambda}\right) u^r\right>_{\phi,t} dz}{\displaystyle\int_{\mathrm{disk}} \left<\left(\rho + \frac{\gamma}{\gamma-1}P_g + b^{\lambda}b_{\lambda}\right) u^t\right>_{\phi,t} dz}
    \ ,
\end{align}
\end{subequations}
where $\langle v^r \rangle$ is the time- and disk-averaged radial velocity weighted by fluid enthalpy.  

The thermal time is estimated using the time-averaged, coordinate-frame radiation flux at the emitting surface, accounting for both radiation diffusion and advection: 
\begin{subequations}
\begin{align}
     t_{\mathrm{thermal}} &= \frac{\displaystyle\int_{\mathrm{disk}} \left<-R^t_{\ t}\right>_{\phi,t} dz}{F_r^{\perp}}
     \ ,
     \\
     F_r^{\perp} &= g_{ij}\left<-R^i_{\ t}\right>_{\phi, t} n_{\mathrm{em}}^j
     \ , 
\end{align}
\end{subequations}
where $F_r^{\perp}$ is the projection of the coordinate-frame radiation flux perpendicular to the emitting surface, and $n_{\mathrm{em}}^j$ is the unit normal vector to that surface.  The emitting surface is defined as either the scattering photosphere or the zero-Bernoulli surface, whichever lies deeper.  In the super-Eddington regime, the scattering photosphere lies above the zero-Bernoulli surface due to the presence of an optically thick wind  (e.g., see E88-a3 in \autoref{fig:profile2d}); in this case, we adopt the zero-Bernoulli surface as the emitting surface.  In the near- and sub-Eddington regimes, the scattering photosphere may lie below the zero-Bernoulli surface (e.g., see E08-a3 in \autoref{fig:profile2d}), and we then adopt the scattering photosphere as the emitting surface.  

The diffusion time is computed similarly to the thermal time, but includes only the contribution from radiation diffusion
\begin{subequations}
\begin{align}
     t_{\mathrm{diff}} &= \frac{\displaystyle\int_{\mathrm{disk}} \left<-R^t_{\ t}\right>_{\phi,t} dz}{\bar{F}_r^{\perp}}
     \ ,
     \\
     \bar{F}_r^{\perp} &= \eta_{ij}\left<-\bar{R}^i_{\ t}\right>_{\phi, t} n_{\mathrm{em}}^j
     \ ,
\end{align}
\end{subequations}
where $\bar{F}_r^{\perp}$ is the fluid-frame radiation flux projected normal to the emitting surface and captures the radiation diffusion process that is responsible for cooling. 

\bibliography{edd_paper_i}
\bibliographystyle{aasjournal}
\end{CJK*}
\end{document}